\documentclass[twoside,twocolumn,english,aps,manuscript]{article}
\usepackage{mathptmx}
\usepackage[T1]{fontenc}
\usepackage[latin9]{inputenc}
\usepackage[a4paper]{geometry}
\usepackage{color}
\usepackage{babel}

\usepackage{float}
\usepackage{url}
\usepackage{amsthm}
\usepackage{graphicx}
\usepackage{setspace}
\usepackage[unicode=true, pdfusetitle,
 bookmarks=true,bookmarksnumbered=true,bookmarksopen=true,bookmarksopenlevel=10,
 breaklinks=true,pdfborder={0 0 1},backref=section,colorlinks=true]
 {hyperref}

\makeatletter

\providecommand{\tabularnewline}{\\}

\newcommand{\lyxaddress}[1]{
\par {\raggedright #1
\vspace{1.4em}
\noindent\par}
}

\@ifundefined{showcaptionsetup}{}{%
 \PassOptionsToPackage{caption=false}{subfig}}
\usepackage{subfig}
\makeatother

\begin{document}

\title{OPUCEM: A Library with Error Checking Mechanism for Computing Oblique
Parameters}

\author{\"Ozg\"ur \c{C}obano\v{g}lu$^{a}$, Erkcan \"Ozcan$^{b}$, Saleh
Sultansoy$^{c}$ and G\"okhan \"Unel$^{d}$ }

\maketitle

\lyxaddress{\noindent $^{a}$European Nuclear Research Center, CERN, Geneva,
Switzerland. $^{b}$Department of Physics and Astronomy, University
College London, London, UK\textcolor{black}{. $^{c}$TOBB ETU, Physics
Department, Ankara, Turkey and Institute of Physics, Academy of Sciences,
Baku, Azerbaijan. $^{d}$Physics and Astronomy Department, University
of California at Irvine, Irvine, USA.}}
\begin{abstract}
After a brief review of the electroweak radiative corrections to gauge-boson
self-energies, otherwise known as the direct and oblique corrections,
a tool for calculation of the oblique parameters is presented. This
tool, named OPUCEM, brings together formulas from multiple physics
models and provides an error-checking machinery to improve reliability
of numerical results. It also sets a novel example for an {}``open-formula''
concept, which is an attempt to improve the reliability and reproducibility
of computations in scientific publications by encouraging the authors
to open-source their numerical calculation programs. Finally, we demonstrate
the use of OPUCEM in two detailed case studies related to the fourth
Standard Model family. The first is a generic fourth family study
to find relations between the parameters compatible with the EW precision
data and the second is the particular study of the Flavor Democracy
predictions for both Dirac and Majorana-type neutrinos.
\end{abstract}

\section{Introduction}

The categorization of the electroweak (EW) corrections based on their
contribution types dates back to a study of photon propagated four-fermion
processes \cite{stu-orig}. The corrections to vertices, box diagrams
and bremsstrahlung diagrams were all considered as {}``Direct''
whereas the propagator corrections due to vacuum polarization effects
were all named as {}``Oblique'' since these participate to the computations
in an indirect manner \cite{stu-def}.

The EW precision data collected over the last few decades by various
particle physics experiments have often been used to constrain many
new models of particle interactions. They are particularly useful
in checking the allowed parameter space of a given model through its
contributions especially to the vacuum polarization corrections to
the boson propagators. The main oblique parameters are usually denoted
by letters $S$, $T$, $U$ and the auxiliaries with letters $V$,
$W$, $Y$ \cite{STU_VWX}. As an example, the $S$ parameter estimates
the size of the new fermion sector and the $T$ parameter measures
the isospin symmetry violation, i.e. the split between the masses
of the new up and down-type fermions. The Standard Model is defined
by the values $S=T=U=0$ for a given top quark and Higgs boson mass.

Together with the detailed review by Peskin and Takeuchi \cite{stu-def},
a number of papers were published, calculating the contribution of
a given model to the oblique parameters. To name a few, the estimation
of the number of fermion families and neutral gauge bosons \cite{numF_Z},
the validity consideration of the Higgs-less models \cite{higssless_STU},
and the investigation of the Majorana nature of the neutrinos \cite{kniehl-kohrs}
can be cited. However, a number of such publications suffer from unusual
notations with typos in formulas, and errors arising due to utilization
of approximations instead of exact calculations (with assumptions
such as $m_{H}>>m_{Z}$ ) or in some cases from unguarded remarks
such as {}``heavy'' for the new fermions.

The goal of this work is two fold: the first is to present a library
to compute the oblique parameters $S$, $T$ and $U$ both with exact
one-loop calculations and with some well-defined approximations for
a number of models and the second is to scan the available parameter
space for the fourth family models. The comparisons between exact
and approximate computations, and amongst formulas from different
papers provide an error checking machinery which improves the enduser
reliability. Implemented in C/C++ languages, we call this library
OPUCEM, which stands for Oblique Parameters Using C with Error-checking
Machinery. The OPUCEM package consisting of the library and a set
of example driver and presentation functions, which are discussed
in this manuscript, are publicly available \cite{opucemWeb}.

The next section describes the technical details of the library implementation.
Then the following sections are on detailed physics studies demonstrating
the use cases of OPUCEM. Section 3 uses the OPUCEM library to investigate
the plausibiliy of a generic fourth Standard Model (SM) family (SM4).
Section 4 focuses on the implications of the EW data from the viewpoint
of the flavor democracy (FD) hypothesis, which provides a principle
theoretical motivation for the potential existence of a fourth SM
family. These case studies deal with defining the parameter regions
favored by the data and define a set of benchmarking points for the
SM4. In both sections, Dirac and Majorana cases of the fourth SM family
neutrino are investigated separately. Finally in Section 5, we present
our concluding remarks.

\section{Function Library Implementation}

The OPUCEM library mainly consists of a header file \texttt{opucem.h}
declaring the available function prototypes and \texttt{opucem.c}
implementing them. The functions are grouped by the relevant physics
cases such as Majorana neutrinos or Higgs bosons. In each case, internal
comments are used to document the source code, indicating the reference
paper for each of the formula and the nature of the calculations (e.g.
exact 1-loop calculations, or approximations valid under certain assumptions
such as the new fermions being much heavier than the Z-boson). Compilation
of the library is straightforward, however a makefile is provided
as is customary. The makefile also features additional targets to
produce example command-line and graphical-user-interface applications
that make use of the library in studying the fermions of a fourth
SM family \cite{6li,WS 1,WS 2}. 

One of the primary goals of the implementation is portability, since
we consider portability an auxiliary measure of the reliability of
the code. To facilitate this goal, the initial implementation was
done in pure C, with the code having no dependencies or requirements
beyond a standards-compliant C compiler. Since certain formulas make
use of complex numbers, we made use of the complex type found in the
C99 standard. However, starting with version 00-00-03, the C99 complex
type was dropped in favor of the std::complex template in the C++
standard template library. The main motivation for this change was
to provide an easy roadmap for an eventual replacement of various
long double variables with multi-precision variables.

The compilation and execution is compatible with any UNIX-like environment
with modern versions of the GNU compiler collection (gcc) \cite{gcc},
that support a subset of the complex functions given in the C99 standard.
For example, the last version which uses the pure C implementation
(v00-00-02) is known to compile with gcc 4.x. On the other hand, the
current version (v00-00-04), which makes use of the C++ std::complex
templates, can only be compiled with versions of g++ 4.2 onwards,
which implements the part of the C++ Technical Report 1 (TR1) that
pertains to complex numbers. TR1 introduced the complex inverse hyperbolic
functions of C99 into the std::complex template, and these additions
are expected to appear in the upcoming C++0x standard as well. 

While we do not intend to support Microsoft Windows operating system
as a regular OPUCEM target, we still have ported the C version of
the library using the GNU development toolchain provided under the
Cygwin environment and independently by the native tools from the
MinGW project \cite{cygwin,mingw}. The main missing ingredients are
complex numbers with long double precision. The latest MinGW distribution
(MinGW 5.1.6) comes with complex.h implementations for gcc 3.4.4 and
gcc 4.3.4 but only with double precision. For this reason, for the
Windows port, we rely on an external implementation of the C99 complex
numbers implemented by S.$~$Moshier \cite{c9x-moshier}. It should
be noted that this complex-number implementation, along with the long
double math library from the same author, makes it possible to port
the C version of OPUCEM to any architecture that supports even rather
old versions of the GNU Compiler Collection. (For example under Cygwin
1.5.25, we were able to compile even with gcc 3.4.4.) Under a separate
Windows branch, we provide documentation and makefiles to facilitate
such efforts.

Finally it should be mentioned that the current version of the code,
while using the C++ complex templates, is still structured in a strictly
C-like manner, making no use of the object-oriented features of C++.
Because of this, it is clear that the enduser experience is likely
to quickly deteriorate as more and more physics models are added to
the library in the future. Therefore for the upcoming revisions of
the code, the planned updates include a better C++ implementation,
in particular making use of namespaces to separate formulas from different
physics models.

The oblique parameter calculation is implemented for the following
scenarios in the current version of the OPUCEM library:
\begin{itemize}
\item New lepton doublets with Dirac \cite{ST_dirac_2hdm} or Majorana \cite{ST_majorana}
type neutral leptons; 
\item New quark doublets with variable $4\times4$ CKM mixing to the third
generation \cite{chanowitz}; 
\item SM and 2 Higgs Doublet Model (2HDM) Higgs bosons \cite{ST_smH,ST_dirac_2hdm}.
\end{itemize}
Additionally, a set of functions for recalculation of the SM origin
in the $S-T$ plane based on reference values of the top quark and
the Higgs boson masses is also available. 

During the implementation of the library, a number of minor errors
on the reference papers have been found and corrected. These are:
\begin{itemize}
\item mixing angle in the definition of the left-handed weak eigenstates
of the two Majorana neutrinos in \cite{ST_majorana}; their equation
10 should read $c_{\theta}^{2}=M_{2}/(M_{1}+M_{2}),\quad s_{\theta}^{2}=M_{1}/(M_{1}+M_{2})\;$.
\item contribution to W and Z boson self energies induced by the two Majorana
neutrinos in \cite{kniehl-kohrs}; the right hand side of their equation
3.6 should start as $...=\frac{{1}}{12\pi^{2}}[[q^{2}-\frac{m_{1}^{2}-m_{2}^{2}}{2}\pm\ldots\;$
(See the corrected form in$~$\cite{daSilva}).
\end{itemize}

\subsection{The error-checking mechanism}

To ensure the correct calculation of the oblique parameters, a number
of validation techniques have been used. The core part of the library
has been pair programmed, with subsequent updates consistently cross-checked
against the earlier parts. Furthermore, a number of sanity checks
have been performed during the implementation. Some of these checks
are documented as embedded comments in the source code. For example,
by setting the masses of the second doublet Higgs particles in the
2-Higgs-Doublet-Model to zero and taking into account the extra multiplicative
factor of 3, we have verified the calculation of the contribution
from the SM Higgs boson. Other sanity checks have been promoted to
become actual run-time tests, accessible through clearly indicated
safe-to-use functions. For example, the contributions to $S$ and
$U$ parameters from Majorana-type neutrinos requires high precision,
for which the long double type might not be adequate under certain
values of the input parameters. To catch such cases automatically,
we check whether the computations for Majorana-type neutrinos converge
to those of the Dirac-type in the fully-degenerate limit of the two
Majorana masses. Moreover, these safe-to-use functions compare approximate
and exact computations and display warnings if absolute or relative
difference between the two are above certain predefined thresholds.
Finally, special cases such as equal mass values for up and down type
fermions, have been addressed carefully with limiting cases of the
formulas derived using symbolic computation tools.

\begin{figure}[!h]
\begin{centering}
\includegraphics[scale=0.24]{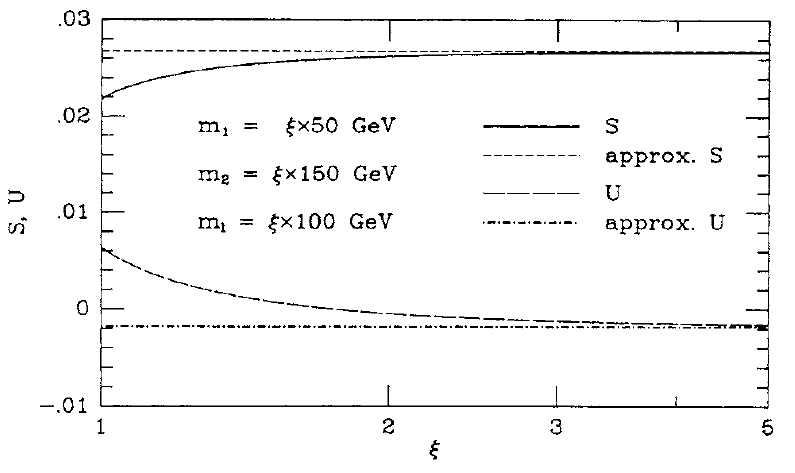}
\par\end{centering}

\begin{centering}
\includegraphics[scale=0.37]{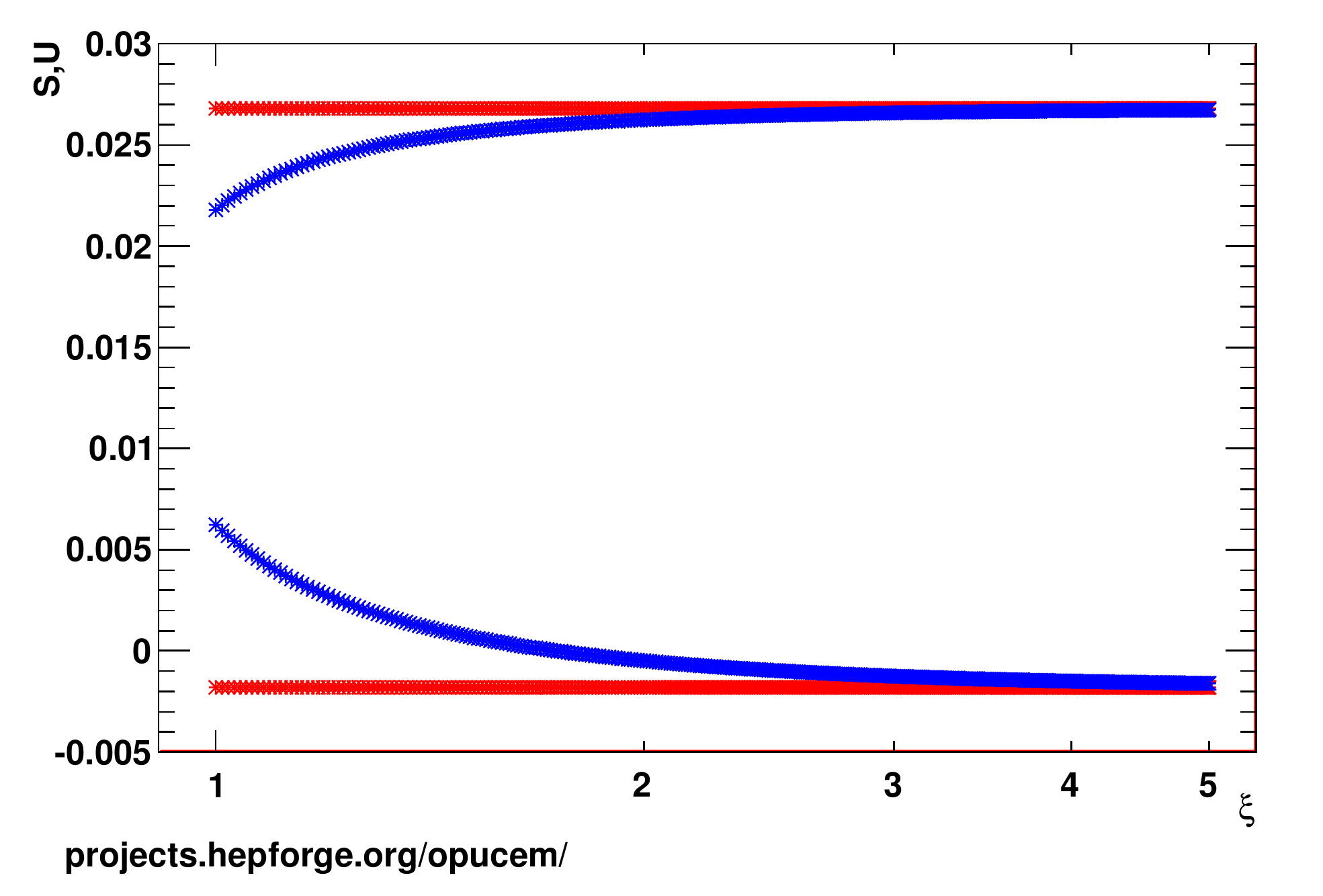}
\par\end{centering}

\caption{Contributions to S and U due to fourth-generation leptons, $N_{1}$,
$N_{2}$ and $\ell$, with masses $m_{1}=\xi\times50$ GeV, $m_{2}=\xi\times150$$\,$GeV
and $m_{\ell}=\xi\times100\,$$\,$GeV as a function of $\xi.$ The
top plot is from \cite{kniehl-kohrs} and the bottom plot is the output
of OPUCEM library. In both cases exact and approximate calculations
are compared.\label{fig:kniehl_kohrs}}

\end{figure}

We have also tried to make sure that the OPUCEM implementation of
the formulas can correctly reproduce the tables and plots in the previously
published work. An example of this has been available in the \texttt{examples/prd48\_pg225}
directory since the very first public version of the library. The
provided shell script generates an executable that calculates $S$
and $U$ as a function of the lepton mass parameter $\xi$, reproducing
the first figure of \cite{kniehl-kohrs}, as shown in Figure$~$\ref{fig:kniehl_kohrs}.

\subsection{Additional tools and presentation functions}

It is worth noting that some of the comparisons mentioned in the previous
section cannot go beyond being qualitative analyses, due to the fact
that quantitative information is missing in the original publications.
In such cases, we have tried to manually extract values from the published
figures and documented the level of agreement as auxiliary notes spread
as comments in the source code. In a few cases, we have also discovered
that the representation of data in the figures was incorrect.

\begin{figure}[!h]
\begin{centering}
\includegraphics[width=0.95\columnwidth]{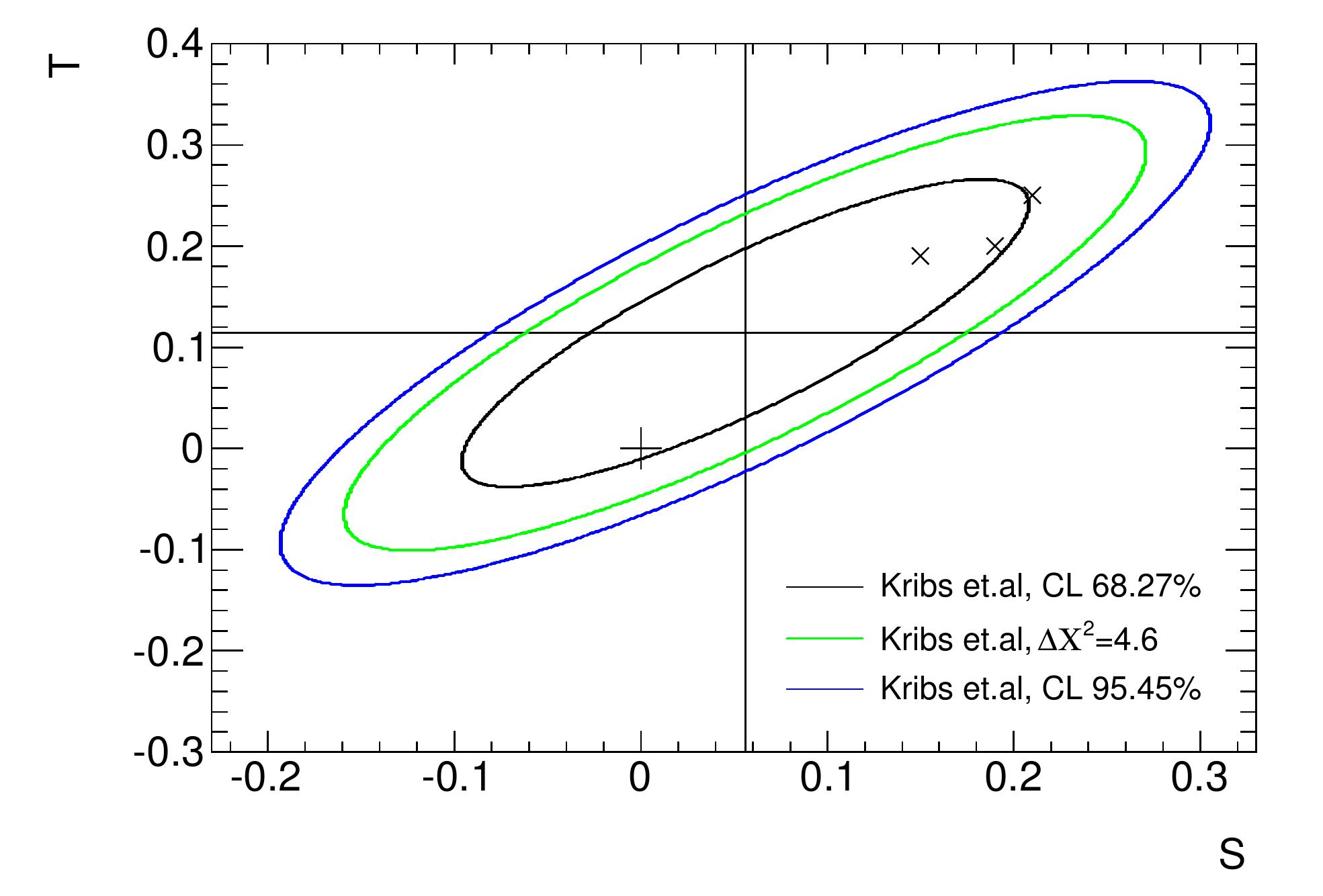}
\par\end{centering}

\caption{Corrected Figure 2 from Kribs et al. \cite{kribbs}\label{fig:Kribs-Corrected-figure}}

\end{figure}

An example of such a mistake is in the Figure 2 of \cite{kribbs},
where the $U=0$ fit results to the EW precision data are being plotted
on the $S-T$ plane. In that figure, the 90\% confidence level (CL)
ellipse seems to have been plotted instead of the 2-sigma (95\% CL)
error ellipse. In our Figure \ref{fig:Kribs-Corrected-figure}, we
show both the 90\% CL and 95\% CL error ellipses to highlight the
enlarged parameter space. In the same paper, we have also identified
additional minor errors in the value of the $T$ parameter for $m_{H}$=300
GeV and in the position of the $S-T$ origin, after their recomputation
to take into account new reference values of the the top quark and
presumed Higgs boson masses. Although the relative size of all these
errors are rather small (\textasciitilde{}3-4\%), and the impact on
the conclusions of the paper is negligible, we mention them here for
the sake of accuracy.

In order to facilitate correct representation of the EW data on the
$S-T$ plane, we provide our plotting routines as part of the OPUCEM
package, under the \texttt{tools/STellipse} directory. Error ellipses
can easily be drawn from the central values and the uncertainties
of the measured $S,\, T$ values and their correlation coefficient.
These plotting functions make use of the ROOT package \cite{root},
an object-oriented C++ analysis framework from CERN and the de facto
standard for statistical analysis in the experimental high-energy
physics community today.

\subsection{Graphical User Interface}

Besides the full-featured standard command-line, the package also
provides users with a light-weight graphical user interface (GUI)
for demonstration purposes. The GUI developed within the ROOT \cite{root}
framework, is actually an independent application loading the OPUCEM
libraries at run-time. It makes relevant function calls based on the
user set of inputs and simply displays the results accordingly on
the $S$-$T$ plane.

\begin{figure}[!h]
\begin{centering}
\includegraphics[width=0.95\columnwidth]{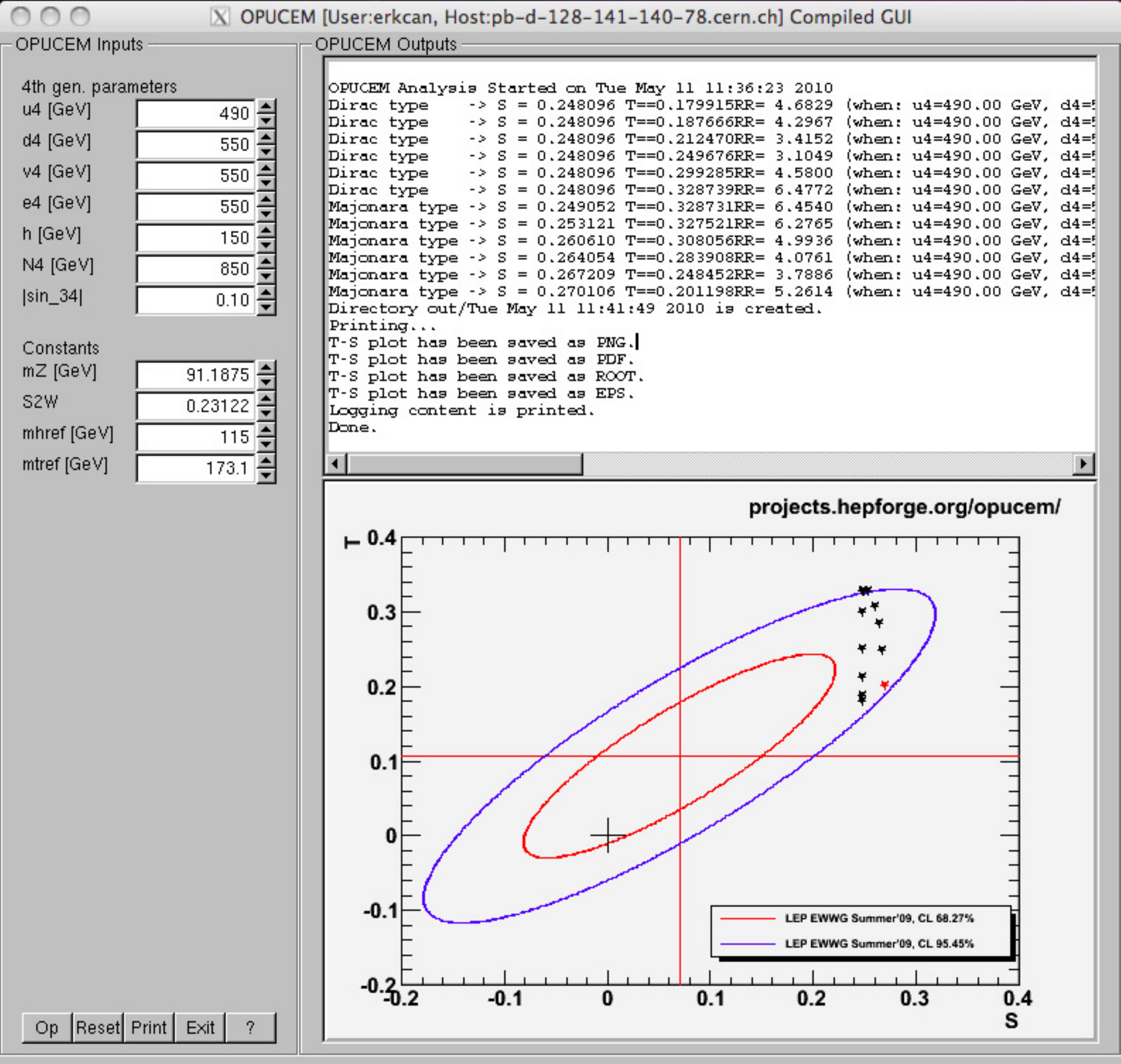}
\par\end{centering}

\caption{The OPUCEM graphical user interface.}
\label{Flo:gui}
\end{figure}

As it can be seen on the Figure \ref{Flo:gui}, the GUI provides simple
input boxes for the input parameters such as the fermion masses and
also access to the internal parameters used in the calculations such
as the sine squared of the weak mixing angle. The user can perform
the $S$-$T$ calculation, see the resulting data points being plotted
and save the plot in multiple formats (e.g. eps, png, root, etc.)
using the buttons at the bottom of the GUI window. For example, majority
of the $S$-$T$ plots shown in this manuscript have been generated
using the OPUCEM GUI.

Like the library itself, the OPUCEM GUI has been tested on various
operating systems (OS) running different ROOT releases. These include
but are not limited to SLC-5, Mac OSX 10.5 and 10.6 running ROOTv5.26.00(b),
Ubuntu-9.10 and 10.4 running ROOTv5.18.00b. However it should be noted
that while the GUI can both be interpreted in ROOT and be natively
compiled, the library functions themselves are always loaded dynamically
in binary form at run time.

We have chosen the ROOT framework for visualization and GUI development
as an example, however we do not necesarily endorse any specific framework.
Since OPUCEM is a generic function library, it can be loaded and instantly
be used from within any rapid application development (RAD) environment
such as MONO\cite{mono} and GAMBAS\cite{gambas} for Basic, QtCreator\cite{qtcreator}
for C/C++, and potentially MS Visual Studio for .NET pretty much the
same way. This makes possible applications of the OPUCEM library available
on almost any platform.

\section{The fourth SM family in General}

For the rest of this paper, we make use of OPUCEM to study the Standard
Model with 4 generations of fermions (SM4) in order to determine the
parameter space compatible with the precision EW data and to look
for possible correlations among the SM4 parameters. A fourth SM family,
with fermion doublets of masses heavier than 100$\,$GeV, could explain
some emerging new physics hints from current particle physics experiments
and also cure some of the theoretical shortcomings of the SM itself
\cite{6li,WS 1,WS 2}. The compatibility of the additional generations
with the electroweak data was previously investigated using a multi-parameter
fit approach$~$\cite{okunlar,chanowitz}.

\subsection{$S,\, T$ Formalism and SM4}

\begin{figure}[!h]
\begin{centering}
\subfloat[]{\includegraphics[bb=110bp 0bp 842bp 595bp,width=0.9\columnwidth]{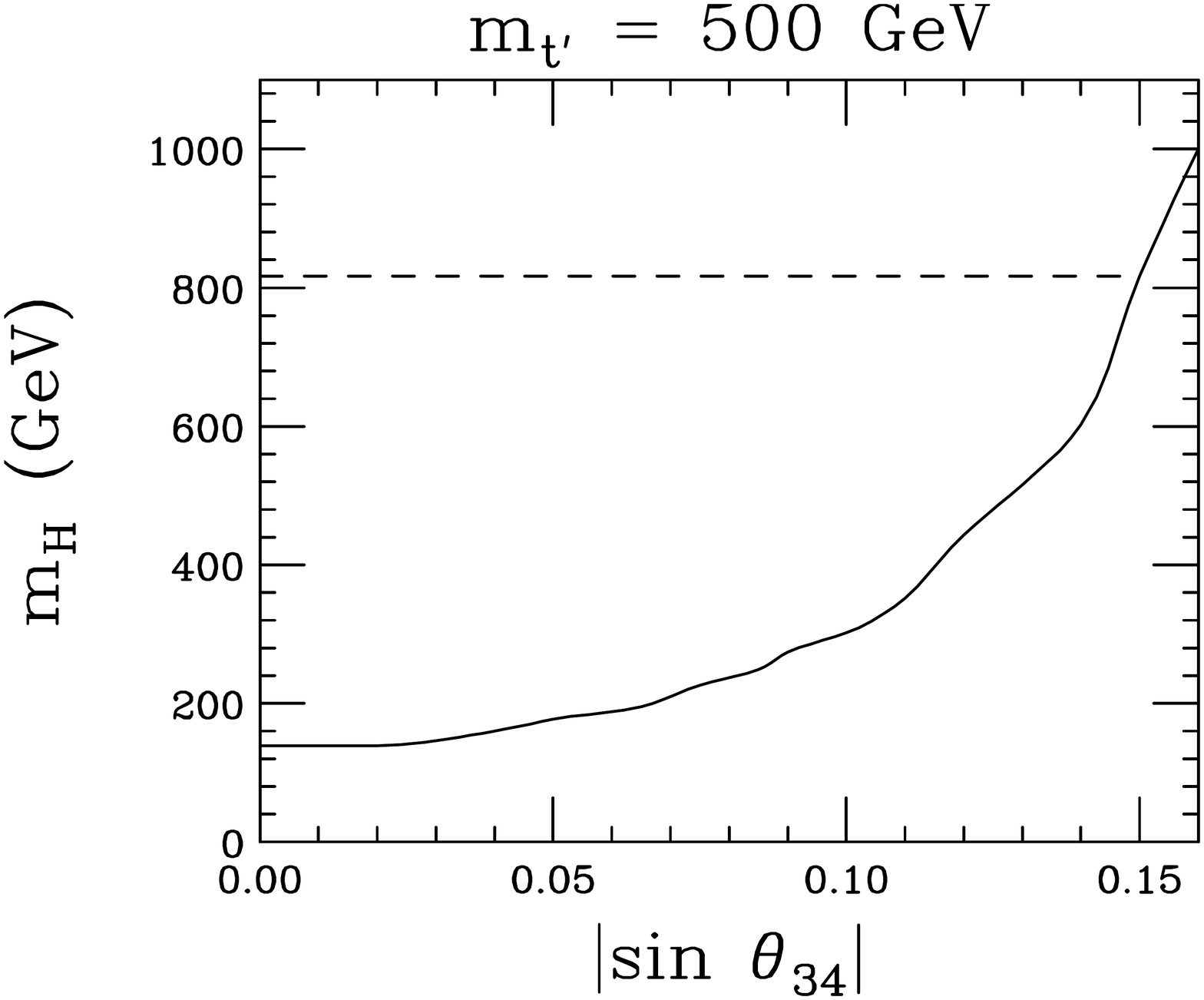}\label{fig:H_vs_sinth34-a}}
\par\end{centering}

\begin{centering}
\subfloat[]{\includegraphics[width=0.9\columnwidth]{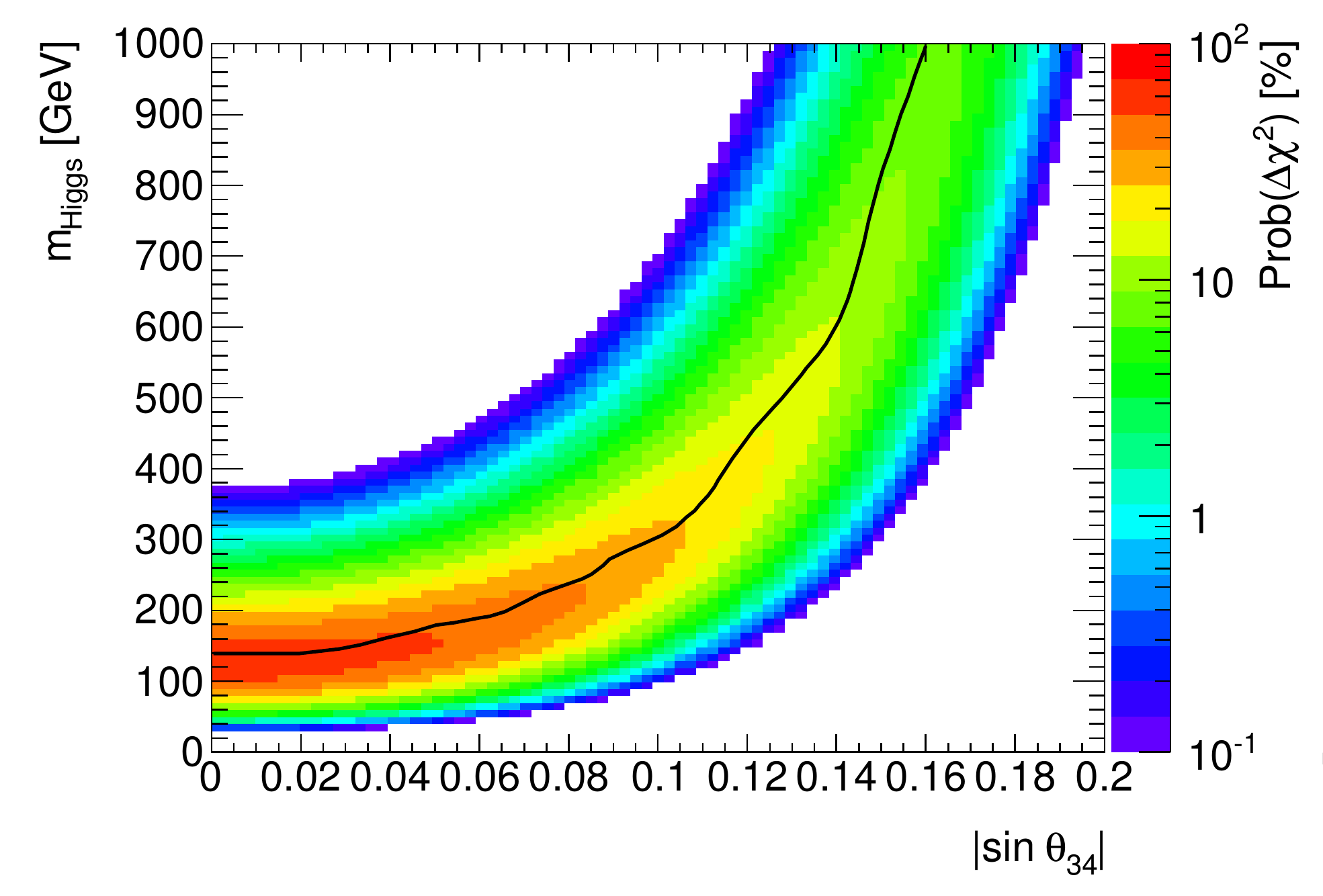}\label{fig:H_vs_sinth34-b}}
\par\end{centering}

\caption{Higgs mass as a function of the mixing angle between 3rd and 4th generations.
(a) Results of the global fit from \cite{chanowitz}. (b) $\Delta\chi^{2}$
probability obtained from a scan of the two parameters as obtained
using OPUCEM. The 4th generation fermion masses are taken to be $m_{e_{4}}=145$$\,$GeV,
$m_{\nu_{4}}=100$$\,$GeV, $m_{u_{4}}=500$$\,$GeV, $m_{d_{4}}=445$$\,$GeV.
The curve from plot$~$(a) has been reproduced on plot$~$(b) for
easy comparison.}

\end{figure}

Before using OPUCEM to scan the fourth generation parameter space,
we would like to point to the excellent applicability of the $S,\, T$
formalism in studies of SM4, by providing an example comparing the
results from a global EW fit (involving 16 input parameters and more
than 10 degrees of freedom) to those obtained from the oblique parameters.
In a recent publication, it has been demonstated that the mixing between
the third and the fourth generation quarks, $\theta_{34}$, plays
an important role in the determination of the best fit to the EW precision
data$~$\cite{chanowitz}. A resulting plot from this publication
is reproduced in Figure \ref{fig:H_vs_sinth34-a}, in which the most
probable value of the Higgs boson mass obtained from the global fit
is plotted as a function of $|\sin\theta_{34}|$. OPUCEM library is
able to closely reproduce this result without a global fit by scanning
the $m_{H}$-$|\sin\theta_{34}|$ parameter space, as shown in Figure
\ref{fig:H_vs_sinth34-b}. The different colors correspond to different
$\Delta\chi^{2}$-probability values. There is a very good agreement
between the most possible values obtained from OPUCEM and the trend
of the global fit results from \cite{chanowitz}. 

\subsection{Scanning SM4 with OPUCEM }

Neglecting possible mixings in the fermionic sector as motivated by
the preference for small $|\sin\theta_{34}|$, one deals with 5 parameters
(masses of the four new fermions, $u_{4},d_{4},e_{4},\nu_{4}$, and
of the Higgs boson) if the neutrino is of Dirac type like the other
fermions. However in the realm of the SM, there is no reason to forbid
the Majorana mass terms for right-handed components of neutrinos.
In this case, the neutrino transforms into two Majorana particles.
We will denote the lighter of these as $\nu_{4}$ and the heavier
as $N_{4}$. Thus in the Majorana case, one has an extra mass parameter
($m_{N_{4}}$) to consider. Given the many degrees of freedom, we
initially follow a simple scanning method: namely, we choose definite
value(s) for Dirac(Majorana) neutrino mass(es) and scan the masses
of the three charged fermions and the Higgs boson. We note that similar
scans with more details have very recently appeared in the literature,
but only for Dirac-type neutrinos$~$\cite{hashimoto,lenz}.

\begin{figure}[!h]
\begin{centering}
\includegraphics[width=0.5\columnwidth]{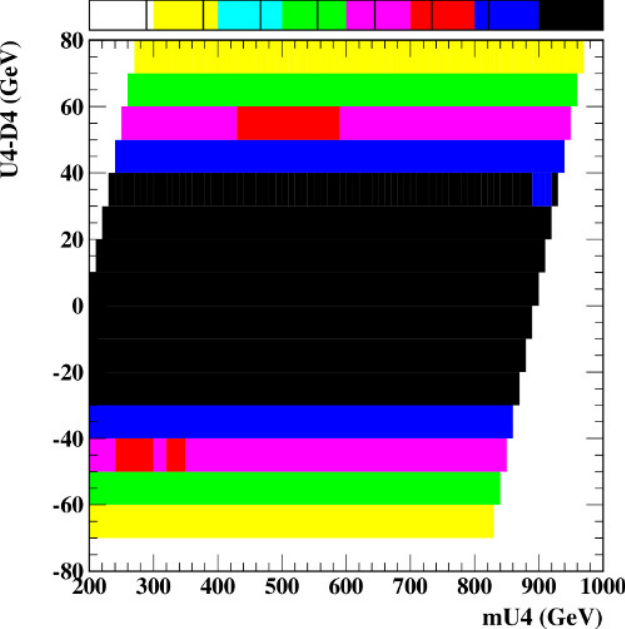}\includegraphics[width=0.5\columnwidth]{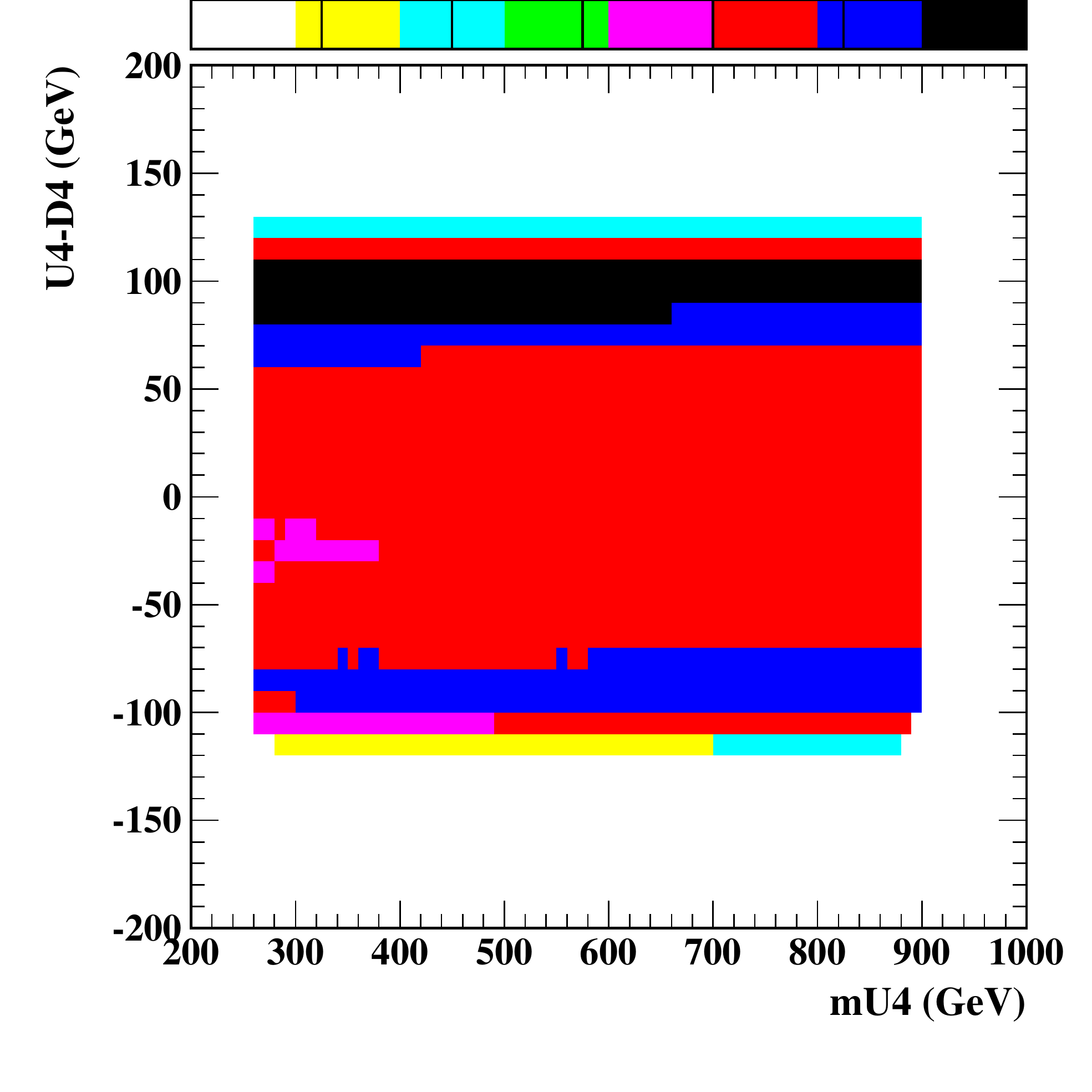}
\par\end{centering}

\caption{The correlation between the fourth family quark masses and mass differences
for Dirac (left) and Majorana (right) cases. \label{fig:genel-quarks}}

\end{figure}

We start by an analysis of the quark sector. The results are presented
in Figure$~$\ref{fig:genel-quarks} for Dirac (left) and Majorana
(right) cases. The mass values, for which the computed \emph{S} and
\emph{T} fall within the 2$\sigma$ error ellipse of the LEP EW working
group's $U=0$ fit$~$\cite{EWWG-2006}, are accepted. Darker regions
correspond to more entries, whereas the white region has no entries.
We scan the charged fermion masses from 200$\,$GeV to 1$\,$TeV and
Higgs mass from 115$\,$GeV to 900$\,$GeV. In the Dirac neutrino
case, the neutrino mass is assumed to be, m$_{\nu_{4}}$= 90$\,$GeV,
the lowest value allowed by the LEP data$~$\cite{PDG}. It is seen
that the preferable values for $u_{4}$-$d_{4}$ mass difference lie
between -30$\,$GeV and +40$\,$GeV, independent of the quark mass.
However, values between -70$\,$GeV and +80$\,$GeV are also allowed.
Studying other neutrino mass values, we observe that these results
show a slight dependence on the fourth neutrino Dirac mass. In the
Majorana neutrino case, we initially assume $m{}_{\nu_{4}}=80$$\,$GeV,
the lowest value allowed by the LEP data$~$\cite{PDG} and $m_{N_{4}}=1000$$\,$GeV.
It is seen that although the preferable values for $u_{4}$-$d_{4}$
mass difference lie around 100$\,$GeV, other values between -110$\,$GeV
and +130$\,$GeV are also allowed. If the lighter neutrino mass increases,
the favoured $u_{4}$- $d_{4}$ mass difference also increases, as
an example for $m_{\nu_{4}}=280$$\,$GeV, the preferred $u_{4}$-
$d_{4}$ mass difference lies around 170$\,$GeV.

\begin{figure}[!h]
\begin{centering}
\includegraphics[width=0.5\columnwidth]{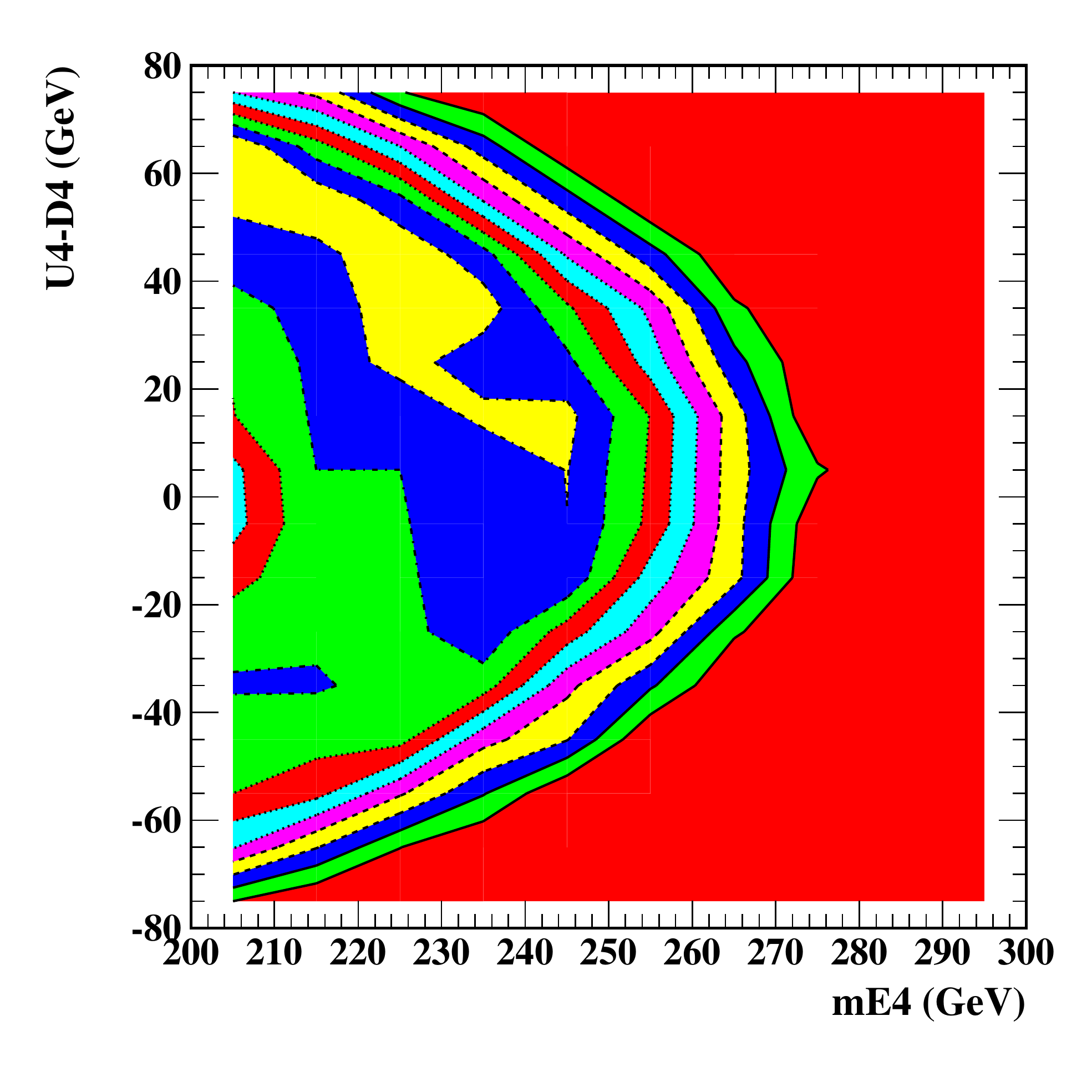}\includegraphics[width=0.5\columnwidth]{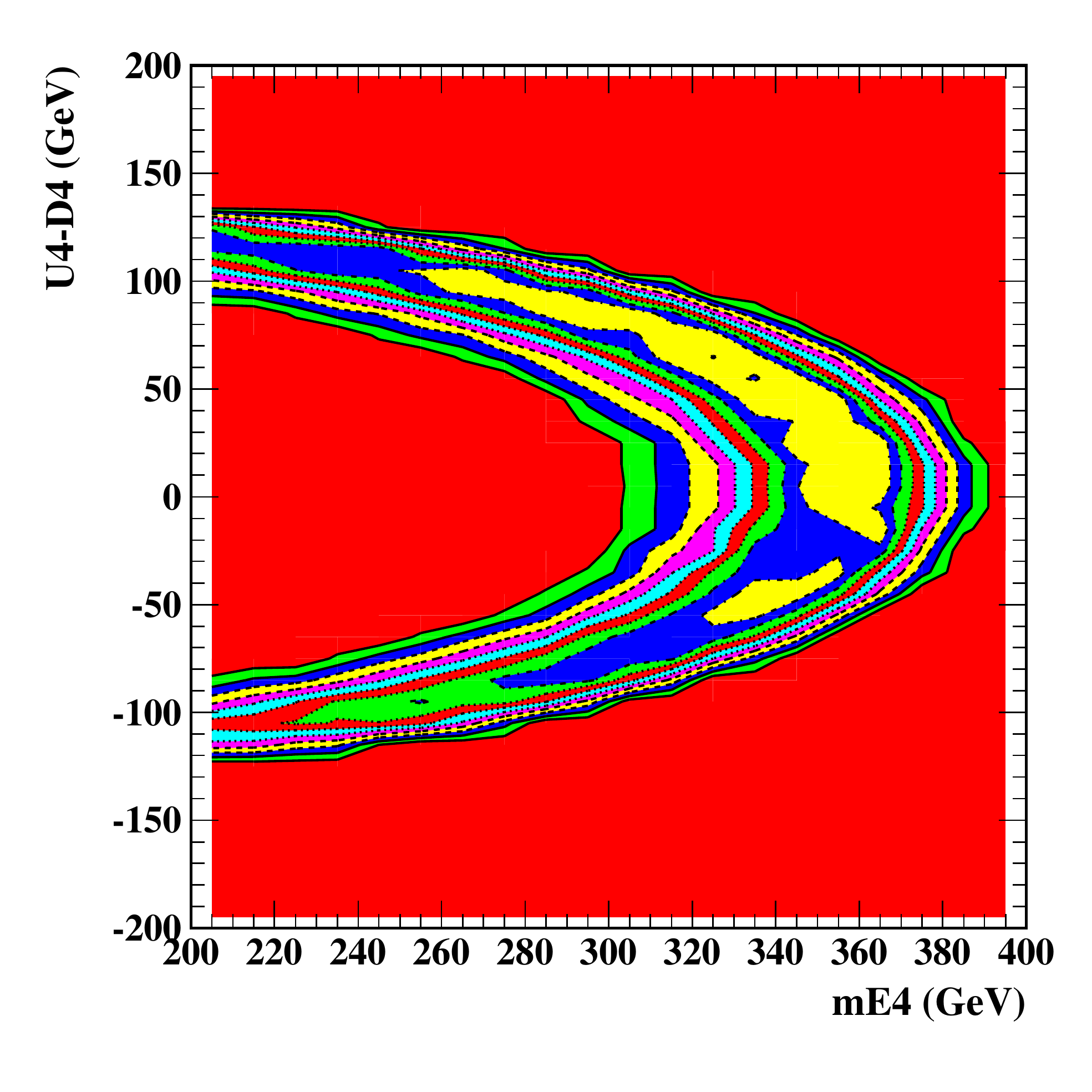}
\par\end{centering}

\caption{The correlation between the fourth family lepton mass and quark mass
differences and for Dirac (left) and Majorana (right) cases. \label{fig:genel-lepton}}

\end{figure}

In Figure$~$\ref{fig:genel-lepton}, the correlation between the
fourth family lepton mass and quark mass differences are shown for
Dirac (left) and Majorana (right) neutrinos. In this figure, where
the darker (lighter) regions correspond to more (less) entries, only
the mass values falling into the 2$\sigma$ error ellipse of \emph{S}
and \emph{T} parameters are shown. The charged fermion masses from
200$\,$GeV to 1000$\,$GeV and Higgs mass from 115$\,$GeV to 900$\,$GeV
have been scanned. Neutrino masses are selected as in Figure$~$\ref{fig:genel-quarks}.

It is seen that only $e_{4}$ masses below 275 (390)$\,$GeV are allowed
for Dirac (Majorana) cases for a $u_{4}$-$d_{4}$ mass difference
that lie between -80$\,$GeV and +90$\,$GeV, independent of the quark
mass.

\subsection{Higgs Mass Dependence}

\begin{figure}[!h]
\begin{centering}
\includegraphics[scale=0.3]{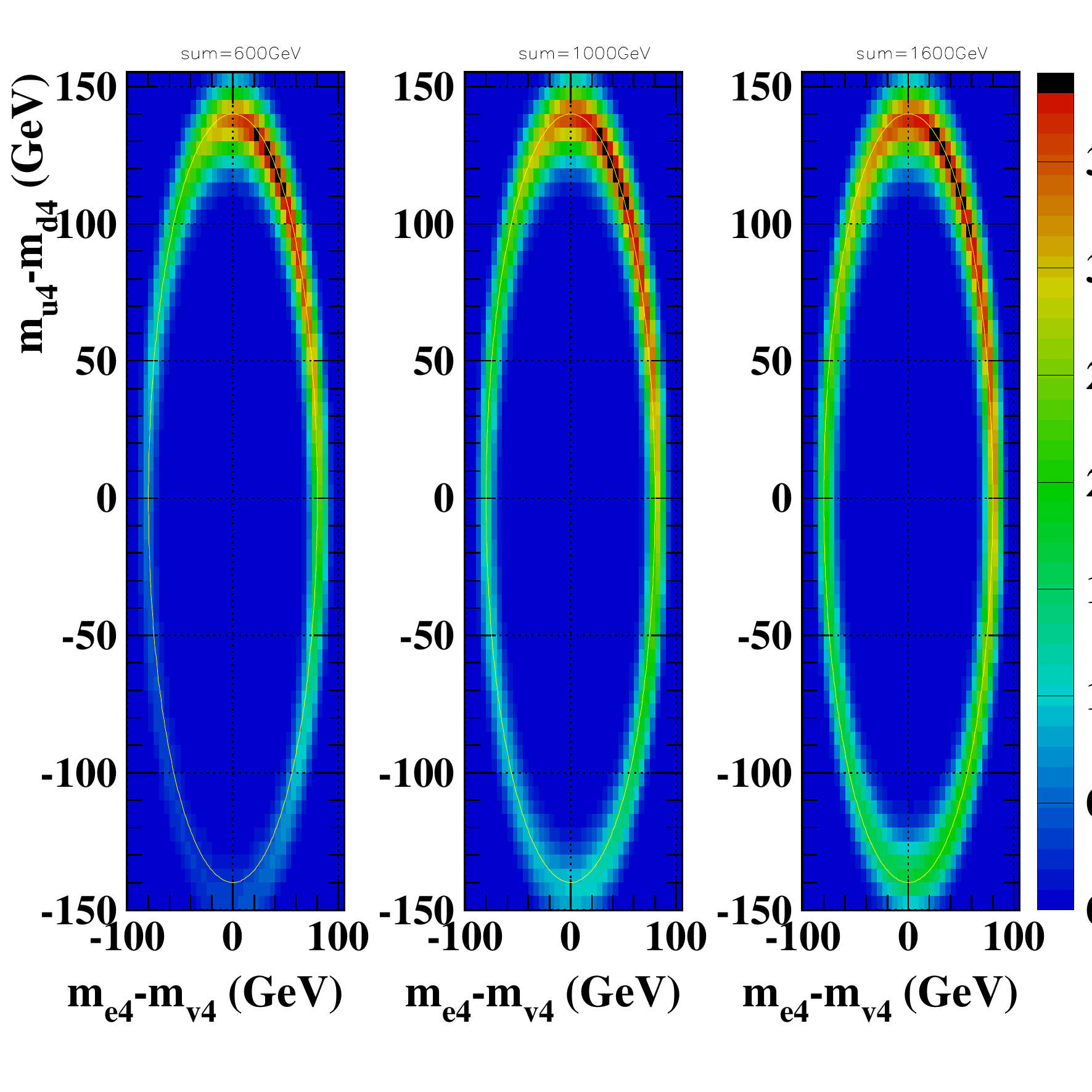}
\par\end{centering}

\begin{centering}
\includegraphics[scale=0.3]{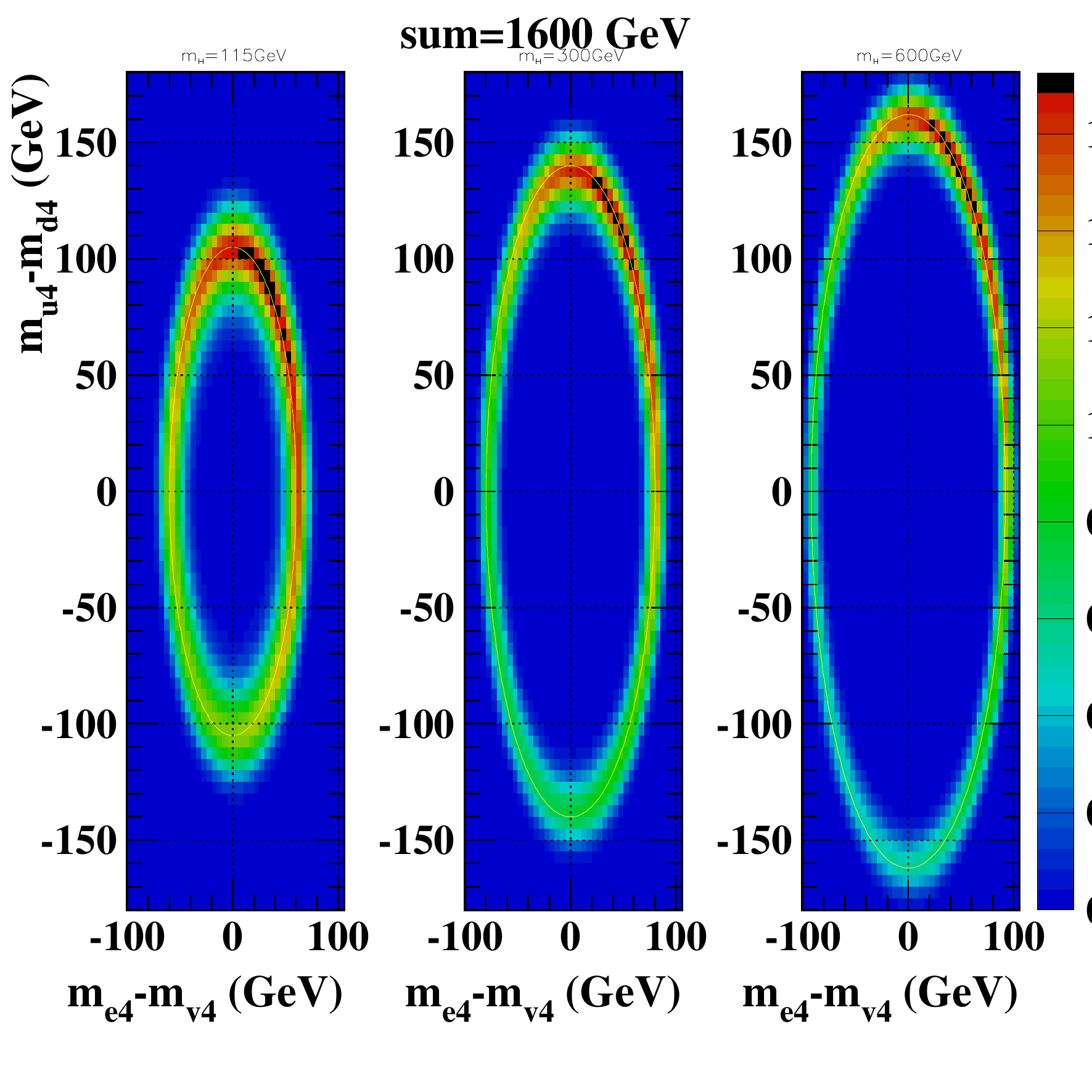} 
\par\end{centering}

\caption{The correlation between the fourth family lepton mass and quark mass
differences for Dirac neutrinos. \label{fig:dirac-sums}}

\end{figure}

The impact of the Higgs mass is better understood with a reduced set
of parameters. To that end, and based on the hint from the mass difference
studies, the correlation of the mass difference between the quark
sector and the lepton sector is studied for different values of the
Higgs boson mass. However, in this study, the fermion masses themselves
are not used as independent variables; while keeping the sum of the
quark and lepton masses as equal and constant, the probabilities are
calculated as a function of $u_{4}-d_{4}$ and $e_{4}-v_{4}$ mass
difference. The upper part of Figure$~$\ref{fig:dirac-sums} shows
the $\chi^{2}$ probability of fourth generation's compatibility with
the EW precision data for $m_{H}=600$$\,$GeV but for different values
of the fermion mass sums: from left to right $m{}_{u_{4}}+m{}_{d_{4}}=m{}_{e_{4}}+m{}_{v_{4}}=$
600, 1000 and 1600$\,$GeV. Different colors correspond to different
probabilities with black being the highest and blue the lowest as
given in the scale on the right hand side. The most probable mass
differences constitute an ellipse in the $m{}_{u_{4}}-m_{d_{4}}$,$m_{e_{4}}-m_{v_{4}}$
plane, independent of the fermion mass sums. The ellipse obtained
from the fit to the 1600$\,$GeV results, can be seen on all 3 plots
in yellow solid line. The impact of the Higgs boson mass is shown
in the lower 3 plots, where the fermion mass sum is taken as 1600$\,$GeV
but the Higgs boson mass changes as 115, 300 and 600$\,$GeV, from
left to right. One can notice that as the Higgs boson mass increases,
so do the fitted ellipse's radii. 

\begin{figure}[!h]
\begin{centering}
\includegraphics[bb=0bp 80bp 567bp 510bp,clip,scale=0.3]{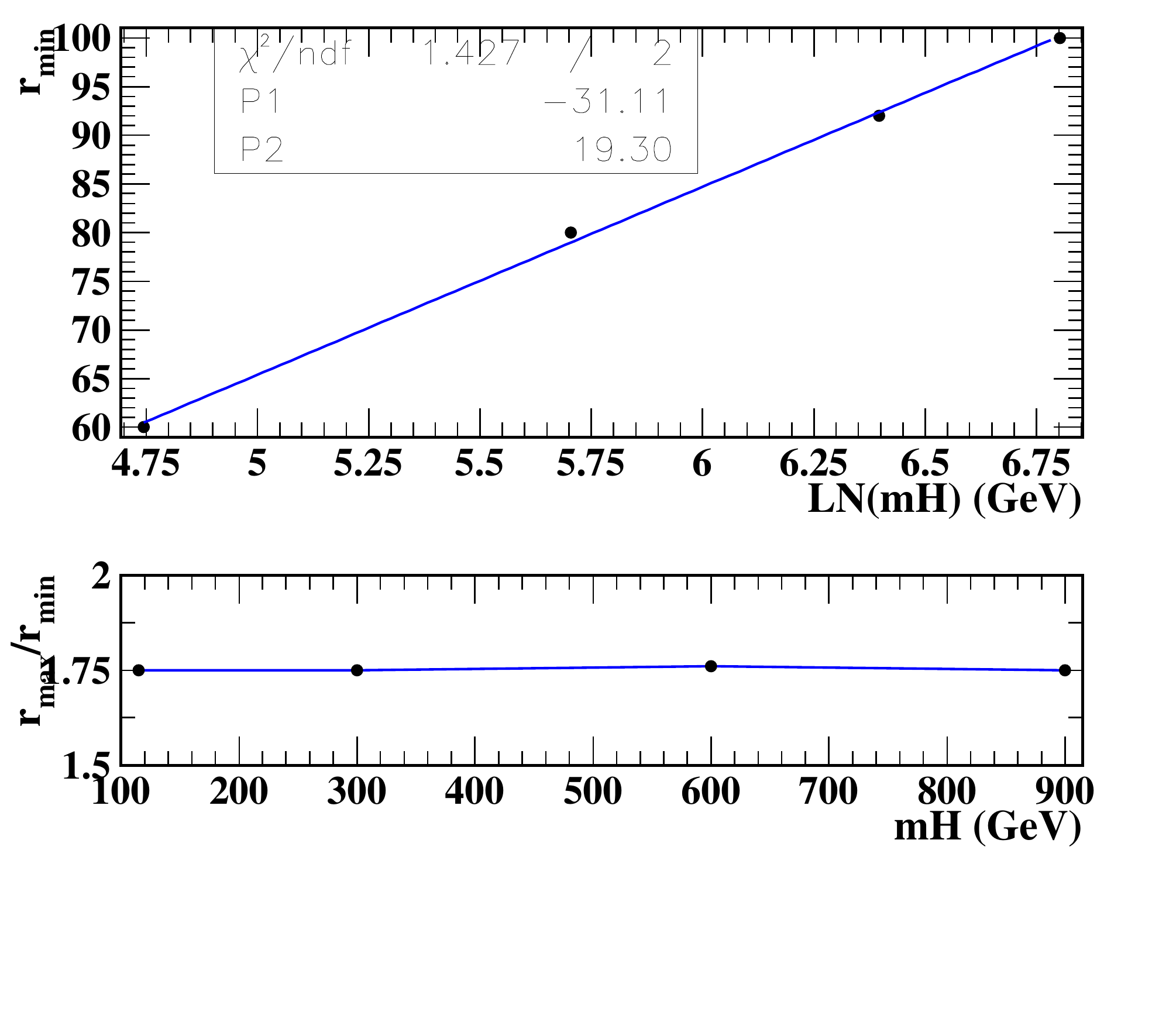} 
\par\end{centering}

\caption{The minor radius and the ratio of the semi-diameters for the most
probable region ellipses as a function of the Higgs mass. \label{fig:fit_sum}}

\end{figure}

The results of the fitted ellipse's semi-minor axis length as a function
of $\ln(m_{H})$ are given in Figure$~$\ref{fig:fit_sum}, the upper
plot. The lower plot shows that the semi-major axis is always 1.75
times longer than its semi-minor axis. It is therefore possible to
summarize the results and parameterize the core of the valid region
as a function of the Higgs mass (all masses in GeV):

\begin{eqnarray}
x & \equiv & m_{e_{4}}-m_{\nu_{4}}\nonumber \\
y & \equiv & m_{u_{4}}-m_{d_{4}}\nonumber \\
r & = & -31.11+19.3\ln(m_{H})\nonumber \\
1 & = & \frac{x^{2}}{r^{2}}+\frac{y^{2}}{(1.75r)^{2}}\end{eqnarray}

\begin{figure}[!h]
\begin{centering}
\includegraphics[width=0.8\columnwidth]{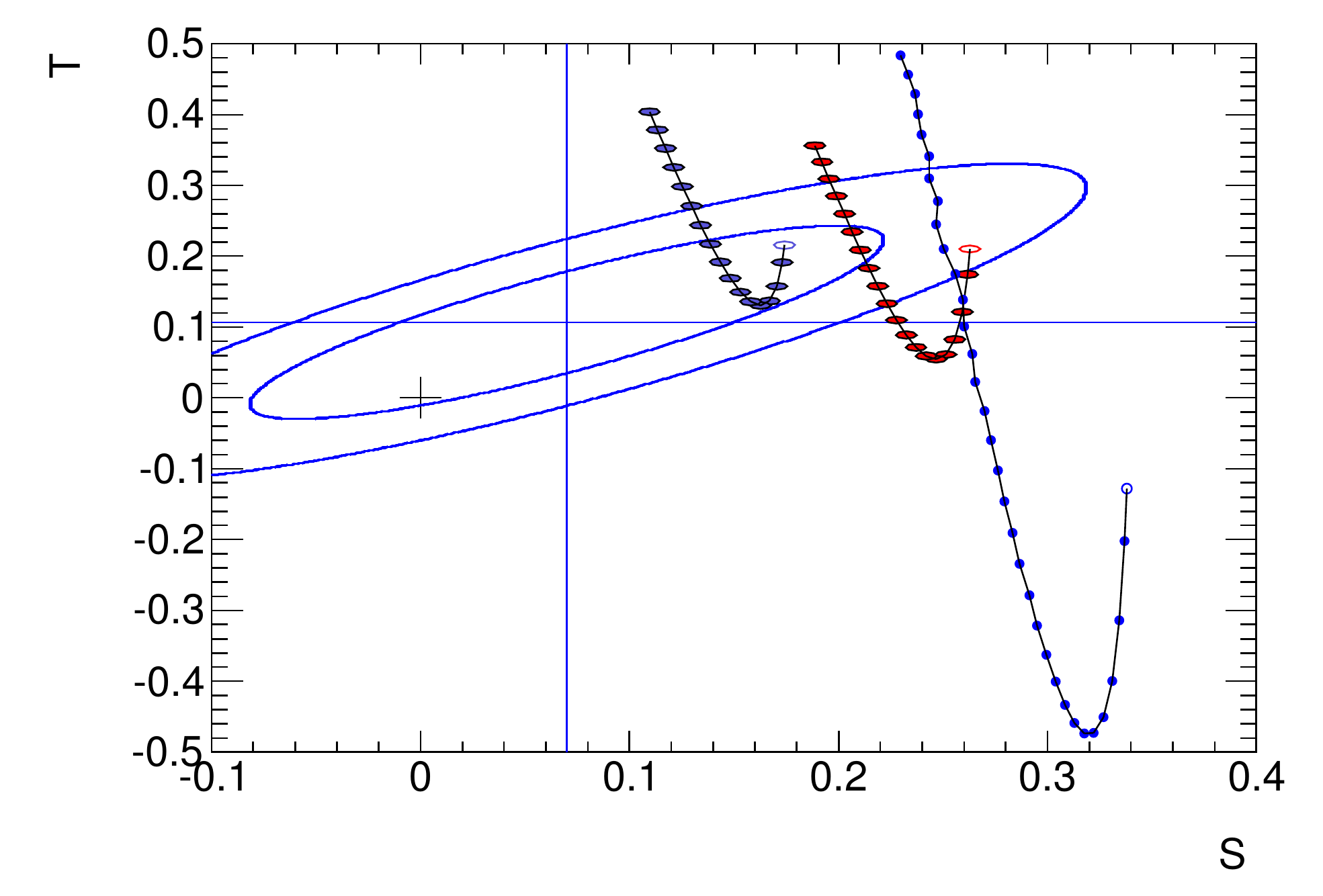}
\par\end{centering}

\caption{The impact of the Higgs boson and the mass ratio of the two Majorana
neutrinos. For three example scenarios with different Higgs boson
masses, the effect of changing the neutrino mass ratio is shown on
the $S$-$T$ plane. In each case, the open circles represent degenerate
Majorana neutrinos (ie. the equivalent of a single Dirac-type neutrino).
The 1 and $2\sigma$ error ellipses are from the 2009 results of the
$U=0$ fit from LEP EWWG$~$\cite{EWWG}.}
\label{Fig:higgs_impact_majorana}
\end{figure}

For Majorana-type neutrinos, the compatibility of the fourth generation
with EW data depends also on the heavier Majorana neutrino's mass.
This dependency can be illustrated using three example scenarios with
different Higgs boson mass values, shown with open circles in Figure$~$\ref{Fig:higgs_impact_majorana}.
The open circle inside the 1-sigma ellipse corresponds to $m_{H}=$115$\,$GeV
and $m_{u_{4}}$=$m_{d_{4}}$=300$\,$GeV, $m_{v_{4}}$=245$\,$GeV
and $m_{e_{4}}$=355$\,$GeV. The open circle inside the 2-sigma ellipse
corresponds to a higher Higgs boson mass, of 450 GeV and to fermion
masses as $m_{u_{4}}$=335$\,$GeV, $m_{d_{4}}$=$m_{v_{4}}$=265$\,$GeV
and $m_{e_{4}}$=335$\,$GeV. The third open circle, the farthest
from the center of the $S$-$T$ minimum is for the heaviest Higgs
mass: 900$\,$GeV with the corresponding fermion masses as $m_{u_{4}}$=$m_{e_{4}}$=435$\,$GeV,
$m_{v_{4}}$=365$\,$GeV and $m_{d_{4}}$=455$\,$GeV. For all the
three open circles the mass of the heavier neutrino, $m_{N4}$, was
initially taken to be the same as its lighter partner. Keeping their
equivalent Dirac mass constant, ($\sqrt{m_{v_{4}}\times m_{N_{4}}}$
= const.), their ratio ($m_{N_{4}}/m_{v_{4}}$) was increased from
1 to higher values in steps of 0.5 alongside the solid line connecting
the closed circles. The highest value of the ratio is fixed by the
lowest experimentally allowed $v_{4}$ mass of 80$\,$GeV. The plot
shows that the increase of the ratio causes an initial decrease in
$T$ which later becomes an increase. This behaviour is accompanied
by a continuous slight decrease in $S$. The combination of these
two has the potential of driving the test points into the allowed
region, even for the $m_{H}$=900$\,$GeV case, the farthest from
the $S$-$T$ minimum. It can be concluded that the asymmetry between
the two Majorana neutrino masses provides an opportunity for accommodating
very heavy Higgs bosons in the Fourth Family model. While the allowed
parameter space for the Dirac case has recently been considered narrow\cite{erler},
it is clear that allowing neutrinos to be of Majorana-type significantly
extends the range of possibilities.

\section{The fourth SM family from Flavor Democracy}

Referring the reader to the original \cite{DMM-0,DMM-1,DMM-2} and
to the review \cite{why the four,FD in particle,naturalness} papers
for details, here we summarize the main FD results. The FD hypothesis
leads the generic case of $n$ SM families to $n-1$ massless families
and one super massive SM family with equal fermion masses. Given the
measured masses of the third SM family fermions, FD requires the existence
of the fourth SM family as the super massive one. Mass and mixing
patterns of the first three family fermions are provided by small
deviations from the full flavor democracy. It should be noted that
FD consequences are independent on concrete mechanism of fermion's
mass generation (e.g. Higgs mechanism), the sole requirement is that
this mechanism should be the same for all SM fermions. 

If the common Yukawa constant of the fermion-Higgs interactions is
taken to be equal to SU$_{W}$(2) coupling constant g$_{W}$, the
fourth family fermion masses are equal to $2\sqrt{2}g_{W}\eta$ (where
$\eta$$\approx$245 GeV), therefore, $m_{4}\approx4\sqrt{2}m_{W}\approx450$
GeV. On the other hand, the fourth family masses are restricted from
above ($m_{4}<1$ TeV) due to partial wave unitarity. Depending on
the nature of the neutrino, again we consider two distinct cases: 
\begin{itemize}
\item The fourth family neutrino is of Dirac nature. Then its mass should
be equal to $m$$_{4}$, leading to $m_{d_{4}}\approx m_{e_{4}}\approx m_{\nu_{4}}\approx m_{4}$.
The Higgs boson's mass, and the size of the quark mixing between the
fourth and the third generations are left as free parameters. 
\item The fourth family neutrino is of Majorana nature. In this case, the
two Majorana particles have masses $m_{\nu_{4}}\approx(m_{4})^{2}/M$
and $m{}_{N_{4}}\approx M$, where $M$ is the Majorana mass scale.
The current experimental lower bound $m$$_{\nu_{4}}>80$$\,$GeV
leads to $M<2500(6100)$$\,$GeV if $m_{4}=450(700)$$\,$GeV.
\end{itemize}
Another possible deviation is due to the large value of the $t$-quark
mass: $u_{4}$ mass is expected to be somehow different than $m_{4}$.
In the remaining of this section we will use OPUCEM to investigate
the compatibility of these estimations with the 2009 EW precision
data \cite{EWWG}.

\subsection{SM4 with Dirac neutrinos}

\begin{figure}[!h]
\begin{centering}
\includegraphics[width=0.8\columnwidth]{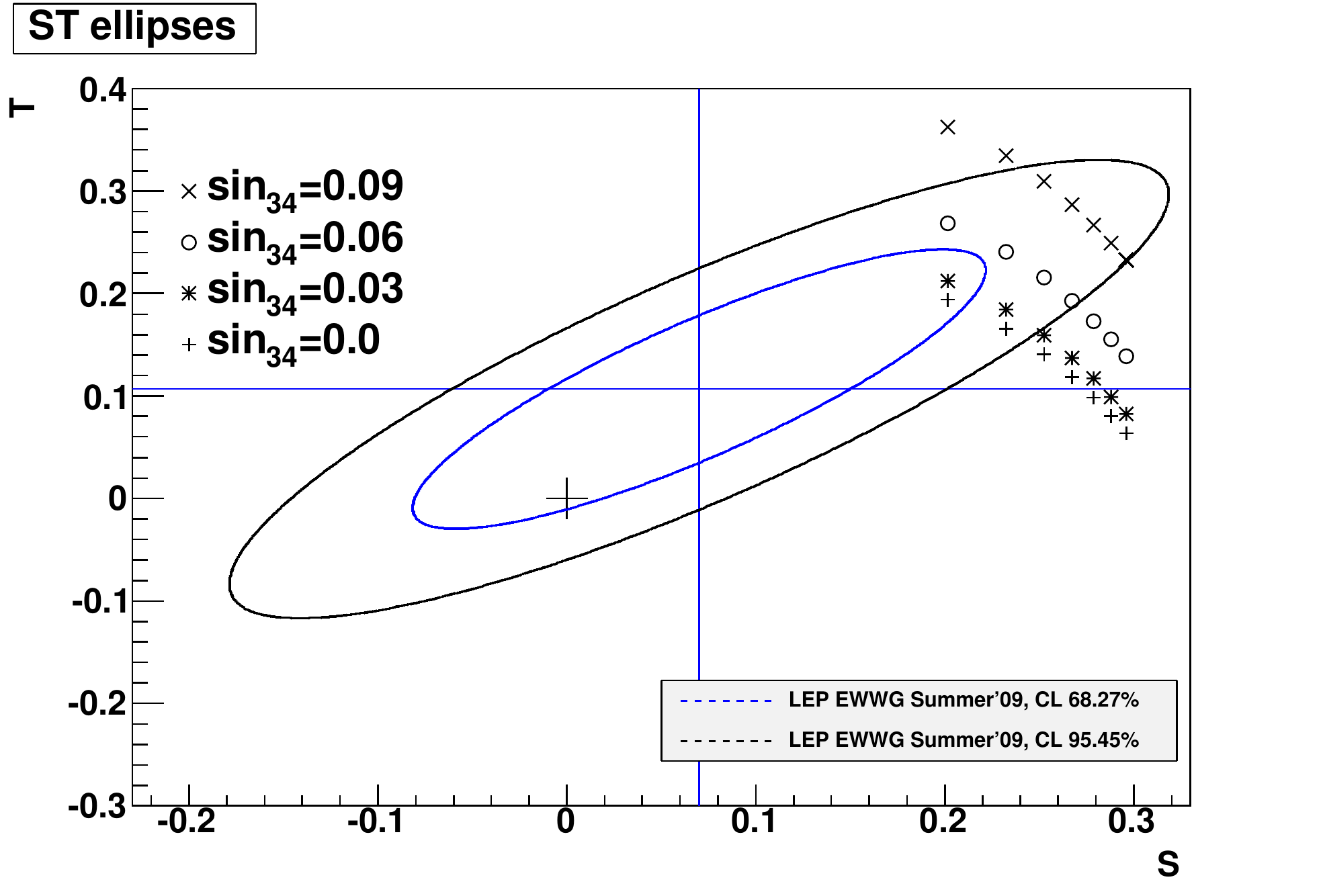}
\par\end{centering}

\caption{The Higgs mass dependence for various values of the third and fourth
family quarks mixing angle, $|\sin\theta_{34}|$. \label{fig:sin34d}}

\end{figure}

For the scenario with Dirac-type neutrinos, we consider an example
case with $m_{d_{4}}=m_{e_{4}}=m_{\nu_{4}}=m_{4}=550$$\,$GeV. The
impact of increasing Higgs boson mass is shown in Figure$~$\ref{fig:sin34d}
for various values of the sine of the mixing angle between the third
and the fourth generations ($\sin\theta_{34}$). The Higgs boson mass
was scanned from 115$\,$GeV up to 415$\,$GeV in steps of 50$\,$GeV.
The $u_{4}$ mass is set to 610 GeV, compatible with the FD predictions
and the findings of the previous section. It can be seen that a larger
mixing angle increases $T$, compensating for larger Higgs boson mass
values pulling towards lower $T$ and slightly higher $S$ values.
The best cases are obtained for either small mixing angle (up to $|\sin\theta_{34}|$
=0.05) and small Higgs mass (up to 150$\,$GeV). Another possibility
resulting in the 2$\sigma$ error ellipse is attained with a larger
mixing angle ( $|\sin\theta_{34}|$ =0.09) and a large Higgs mass
(around 300$\,$GeV).

\begin{figure}[!h]
\begin{centering}
\includegraphics[width=0.8\columnwidth]{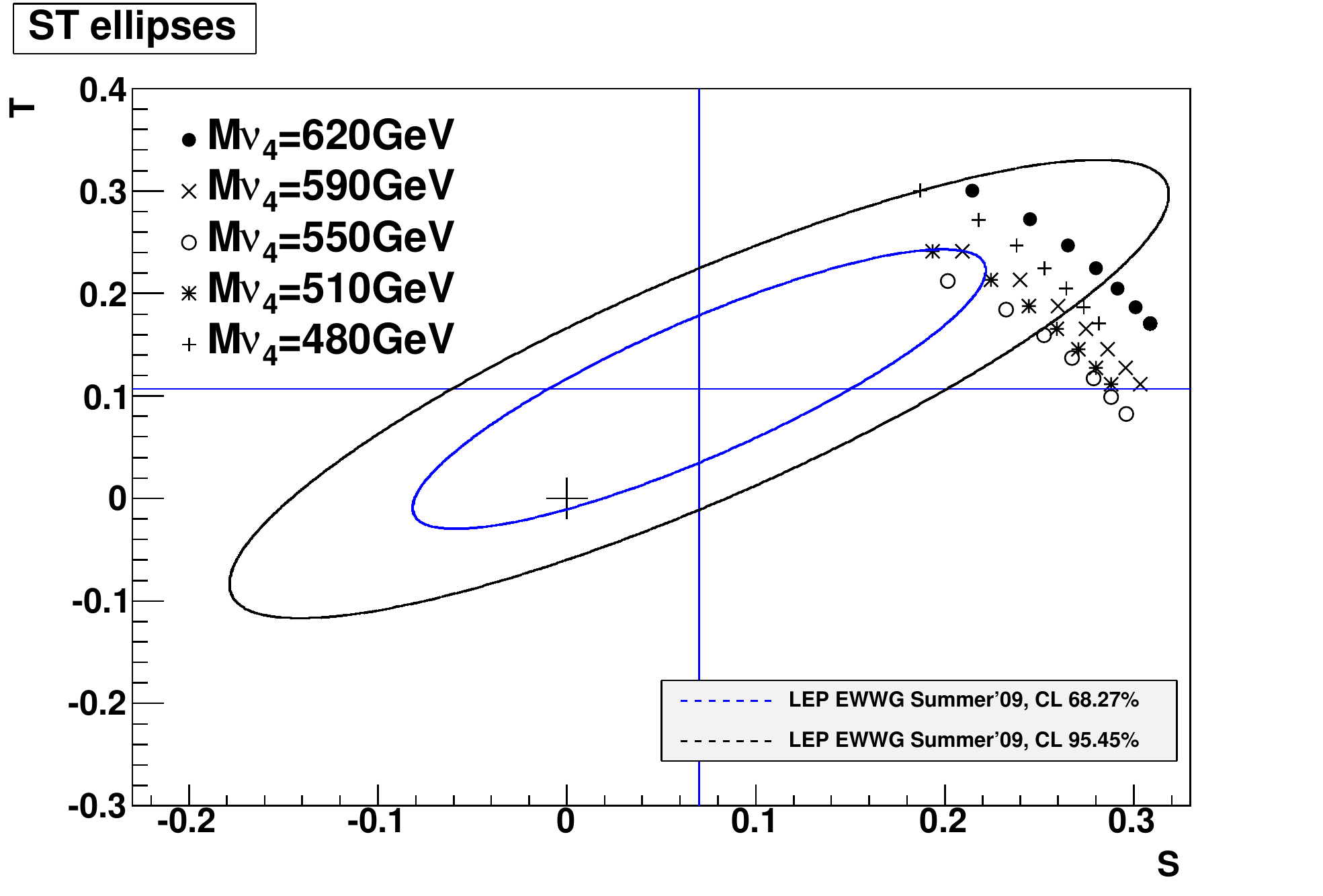}
\par\end{centering}

\caption{The Higgs mass dependence for various values of the fourth family
lighter neutrino, $\nu_{4}$ . \label{fig:mn4d}}

\end{figure}

A similar approach was taken to study the impact of the neutrino mass,
$m_{\nu_{4}}$, on the fourth family model's compatibility with the
EW precision data. The Figure$~$\ref{fig:mn4d} contains a scan of
the Higgs boson mass, from 115 GeV up to 415 GeV in steps of 50 GeV
for 5 different values of $m_{\nu4}$, 480, 510, 550, 590 and 620
GeV. The other relevant parameters are $m_{u_{4}}=610$$\,$GeV and
$m_{d_{4}}=m_{e_{4}}=550$$\,$GeV. The third and fourth family quarks
mixing angle is set as $|\sin\theta_{34}|=0.03$ . One can conclude
that although $m_{\nu_{4}}=550$$\,$GeV and $m_{H}=115$$\,$GeV
describe the parameters most suitable to EW data and also compatible
with FD, a change in the $\nu_{4}$ mass up to $\pm$70$\,$GeV is
also allowed as long as the Higgs boson mass remains less than 215
GeV.

\begin{figure}[!h]
\begin{centering}
\includegraphics[width=0.8\columnwidth]{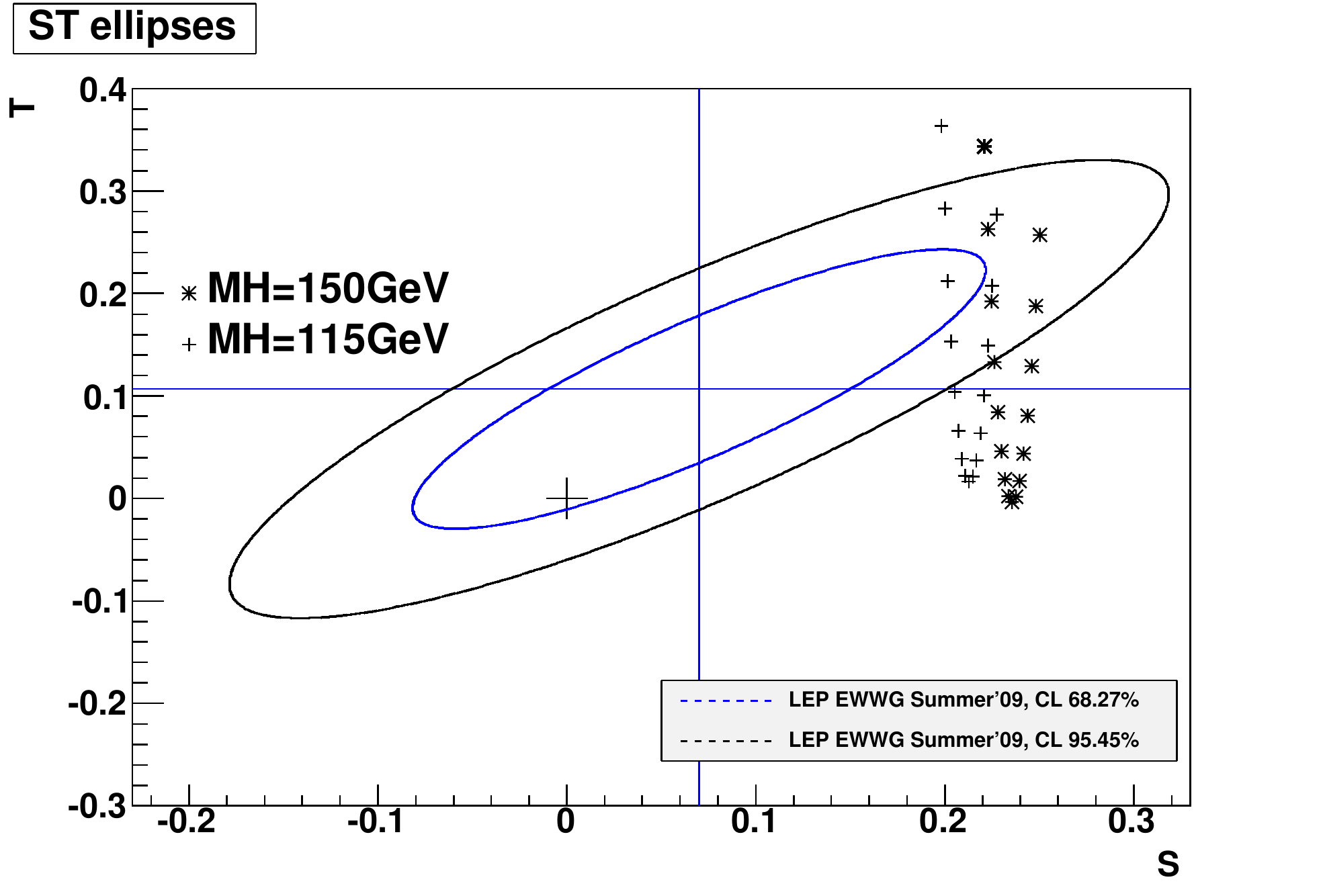}
\par\end{centering}

\caption{The fourth family up type quark, $u_{4}$ mass dependence for various
values of the Higgs mass. \label{fig:MHd}}

\end{figure}

The impact of the change in $u_{4}$ mass is shown in Figure$~$\ref{fig:MHd}
for two Higgs boson mass values. The other relevant parameters are
$m_{d_{4}}=m_{e_{4}}=$$m_{\nu_{4}}$=550 GeV and $|\sin\theta_{34}|=0.03$.
The $u_{4}$ mass was scanned from 480 up to 630$\,$GeV in steps
of 10$\,$GeV. As the $u_{4}$ mass increases, $T$ value decreases
and hits a minimum, after that $T$ keeps increasing, meanwhile $S$
decreases slowly but constantly. For larger values of the Higgs mass
the same behaviour is observed with a starting point consistent with
previous observations, i.e. an increase in $S$ and a decrease in
$T$. One can conclude that although the $m_{H}$=115 GeV and $m_{u_{4}}=610$$\,$GeV
is most compatible with the EW fits, other $u_{4}-H$ mass pairs are
also yielding results very close to 1$\sigma$ ellipse boundary; e.g.
$m_{H}=115$$\,$GeV, $m_{u_{4}}=490$$\,$GeV or $m_{H}=150$$\,$GeV,
$m_{u_{4}}=610$$\,$GeV. It is interesting to note that the $T$
minimum is obtained for $m_{u_{4}}=m_{4}=550$$\,$GeV.

\begin{figure}[!h]
\begin{centering}
\includegraphics[width=0.8\columnwidth]{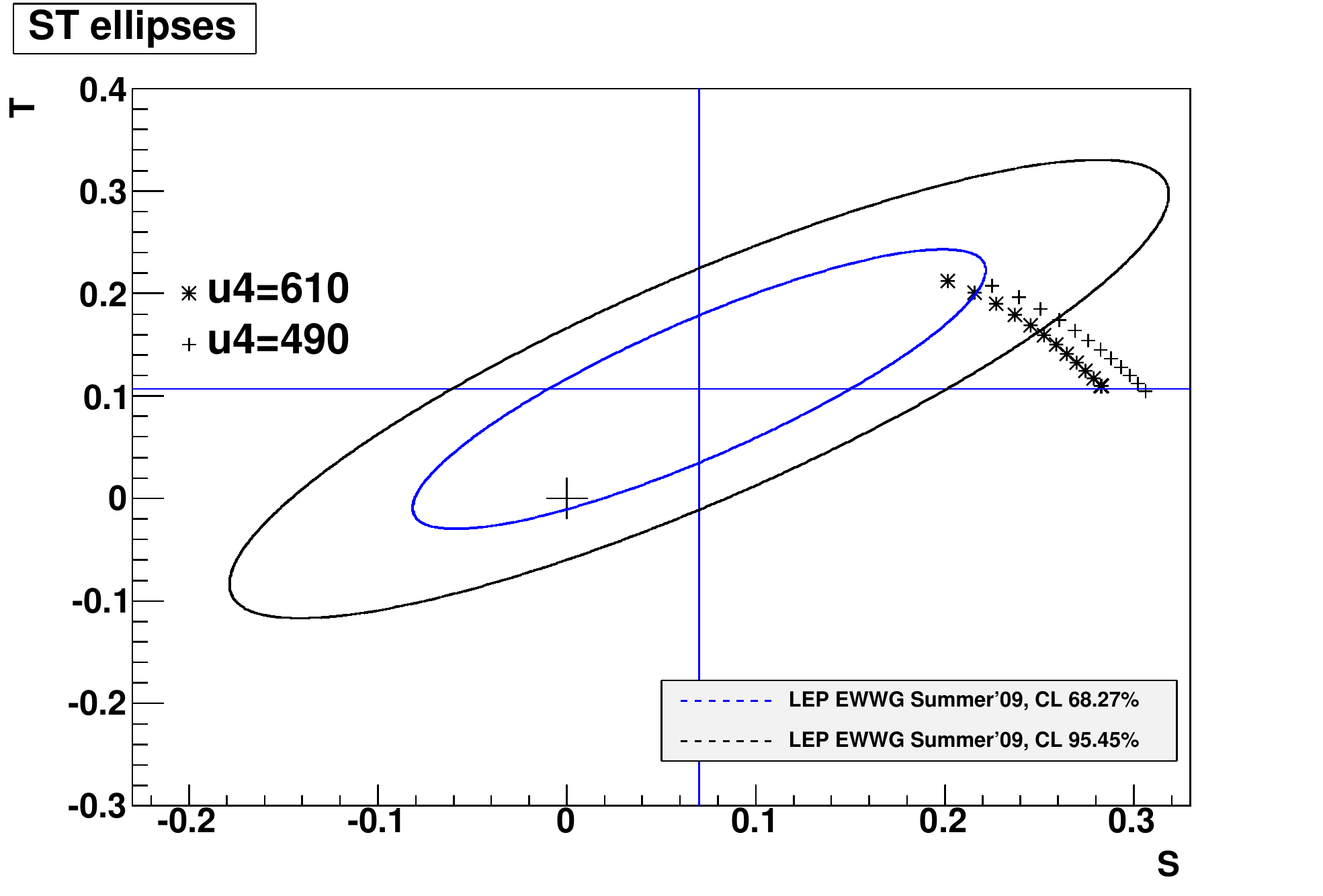}
\par\end{centering}

\caption{The Higgs boson mass dependence for various values of the fourth family
up type quark, $u_{4}$. \label{fig:u4d}}

\end{figure}

A cross check was performed by taking the two most compatible $u_{4}$
mass values found in the previous paragraph and the Higgs mass was
increased from 115 up to 335$\,$GeV in steps of 20$\,$GeV. The other
parameters were kept same as before, namely $m{}_{d_{4}}=m_{e_{4}}=m_{\nu_{4}}=550$$\,$GeV
and $|\sin\theta_{34}|=0.03$ . The results are shown in Figure$~$\ref{fig:u4d}.
One can conclude that for $m{}_{u_{4}}=$610$\,$(490)$\,$GeV Higgs
masses up to 215$\,$(175)$\,$GeV give results within the 2$\sigma$
ellipse. However the smaller Higgs masses are favoured.

\begin{figure}[!h]
\begin{centering}
\includegraphics[width=0.8\columnwidth]{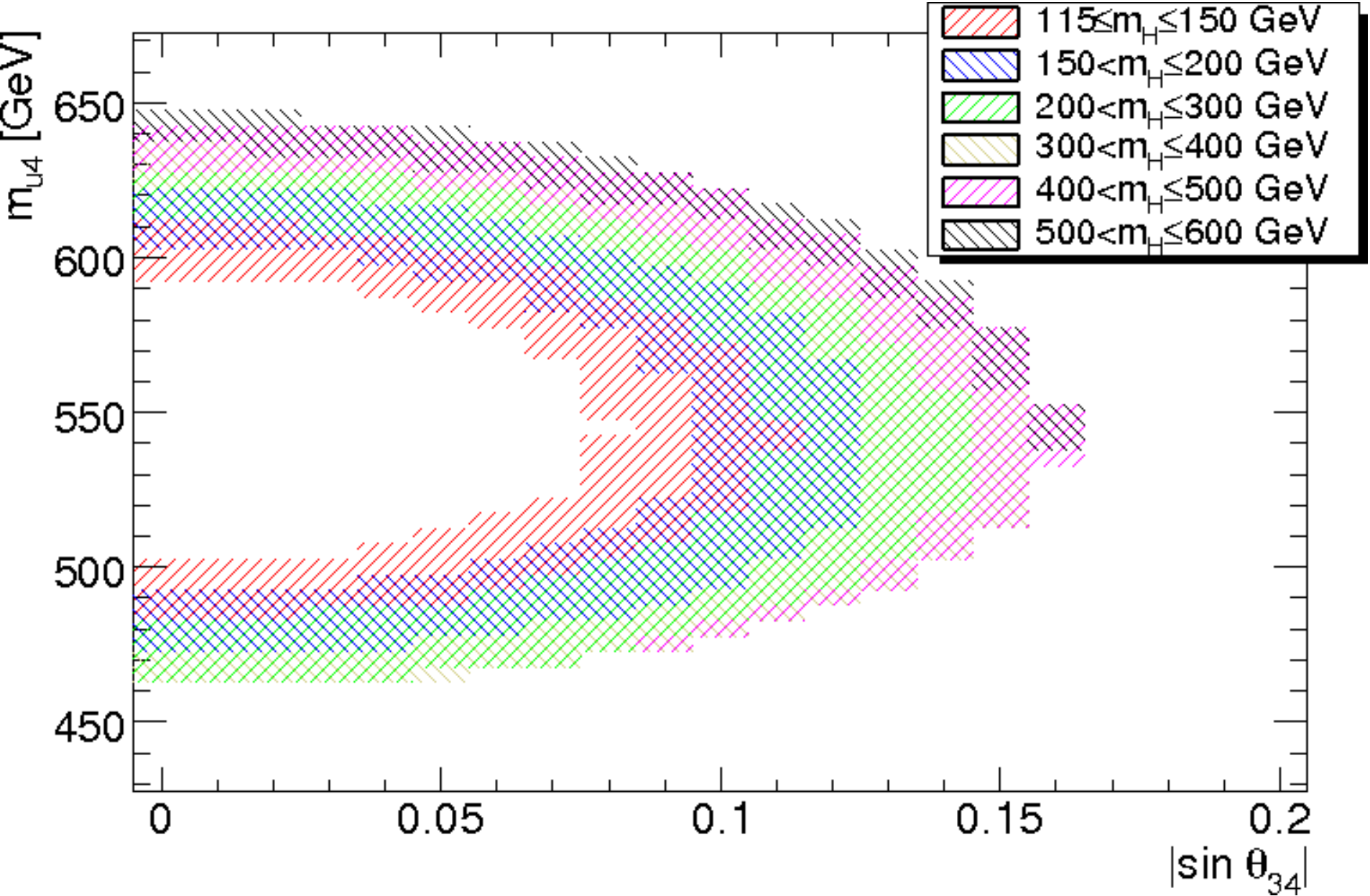}
\par\end{centering}

\caption{The $u_{4}$ vs $|\sin\theta_{34}|$ scan. The contributions from
Higgs bosons with different mass values are shown in different colors.
\label{fig:The-u4-s34} }

\end{figure}

The third and fourth families mixing angle dependence has also been
investigated. The Figure$~$\ref{fig:The-u4-s34}, shows all the mass
points within the $2\sigma$ error ellipse resulting from a scan of
$m_{u_{4}}$ and $|\sin\theta_{34}|$ values. The contributions from
Higgs bosons with different mass values are shown using different
markers. Other relevant parameters are kept as before, namely $m{}_{d_{4}}=m{}_{e_{4}}=m{}_{\nu_{4}}=550$$\,$GeV. 

For $|\sin\theta_{34}|$ up to 0.07, there are two allowed $u_{4}$
mass regions: one above and one below the $m_{4}$ mass. Inside these
regions, smaller (larger) Higgs masses necessitate $u_{4}$ masses
closer (farther) to the $m_{4}$ mass. With the increase of $|\sin\theta_{34}|$
value, the two regions merge at 0.07 and continue up to 0.17, the
maximum allowed value. One should also note that the $m{}_{u_{4}}\geq m_{4}$
case has a larger allowed range of parameters, it allows the highest
values of the Higgs mass, its larger portion survives if the mass
points are required to reside in the $1\sigma$ error ellipse. Therefore
according to FD, and also following the observed hierarchy of the
second and third generations, one expects $m_{u_{4}}>m_{d_{4}}$.

Finally for different values of $m_{4}$, the free parameters of the
Dirac case were scanned. The Table$~$\ref{Flo:Best-R-Dirac} contains
the mass and mixing parameters for the best $\triangle\chi^{2}$ values
obtained from a scan of all relevant parameters. It is seen that the
favoured values are $m_{H}=115$$\,$GeV, $|\sin\theta_{34}|=$0.02
and $m{}_{u_{4}}-m_{4}=60$$\,$GeV, independent from $m_{4}$ mass.
For all three $m_{4}$ values, we find that $\triangle\chi^{2}$ comparable
to or lower than the 3-family SM's $\triangle\chi^{2}=1.7$ is achievable. 

\begin{table}[H]
\caption{The fourth family parameters at the best $\triangle\chi^{2}$ values
compiled for FD hypothesis, Dirac scenario.}

\centering{}\begin{tabular}{c|c|c|c}
$m_{4}$ (GeV) & 400 & 550 & 700\tabularnewline
\hline 
$m_{u4}$ (GeV) & 460 & 610 & 760\tabularnewline
\hline 
$\sin\theta_{34}$ & 0.03 & 0.02 & 0.02\tabularnewline
\hline 
$m_{H}$ (GeV) & 115 & 115 & 115\tabularnewline
\hline 
$\triangle\chi_{min}^{2}$ & 1.58 & 1.67 & 1.72\tabularnewline
\hline 
$S$ & 0.198 & 0.202 & 0.204\tabularnewline
\hline 
$T$ & 0.202 & 0.202 & 0.208\tabularnewline
\end{tabular}\label{Flo:Best-R-Dirac}
\end{table}

\subsection{SM4 with Majorana neutrinos}

\begin{figure}[H]
\begin{centering}
\includegraphics[width=0.8\columnwidth]{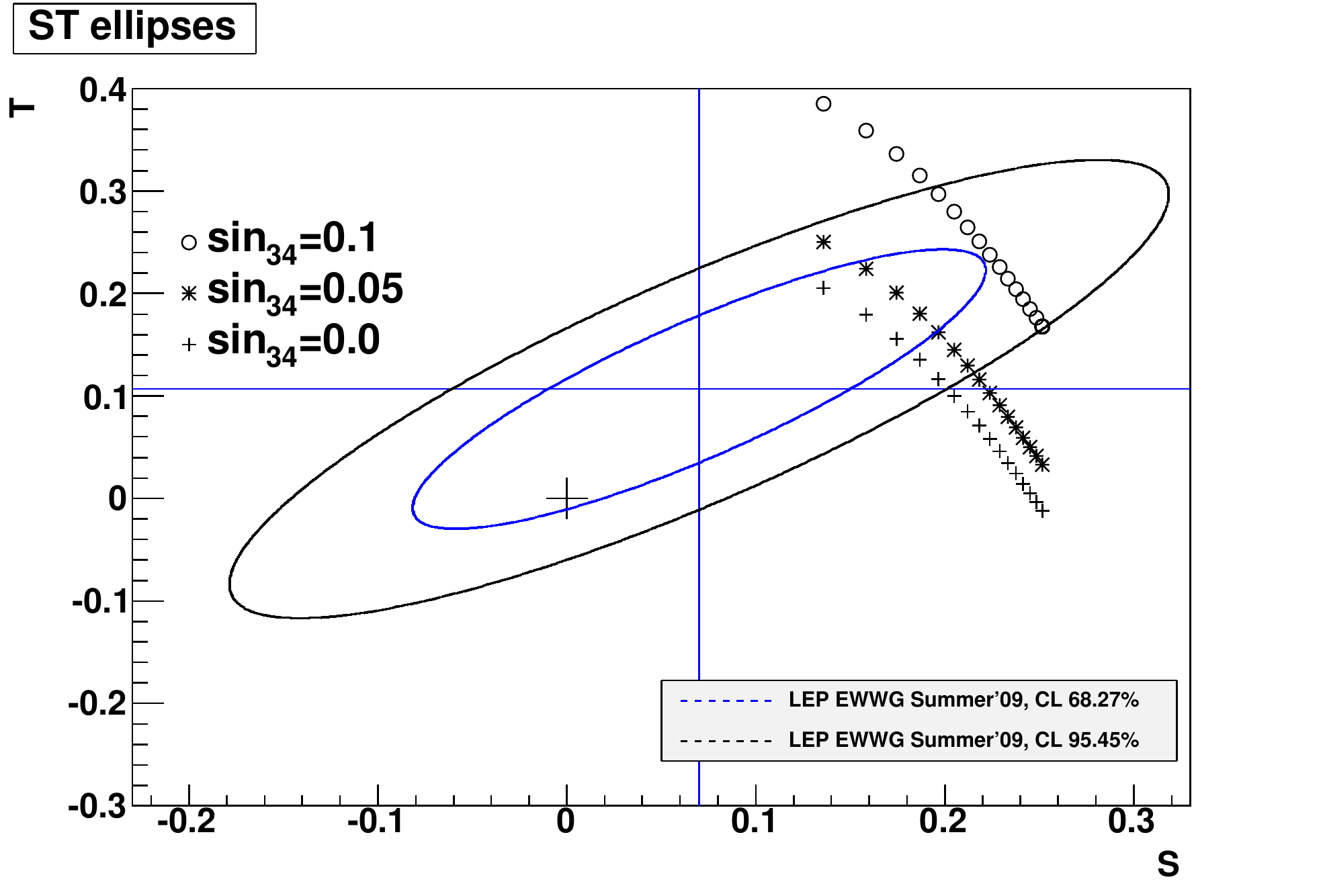}
\par\end{centering}

\caption{The Higgs mass dependence for various values of the third and fourth
family quarks mixing angle, $|\sin\theta_{34}|$ . \label{fig:sin34m}}

\end{figure}

If the neutrinos are of Majorana type, an additional parameter, namely
the mass of the heavier neutrino ($m_{N_{4}}$) has to be taken into
account. This section deals with this case using an approach similar
to the previous section. The default mass pattern is set as: $m_{d_{4}}=m_{e_{4}}=m_{4}=550,\, m_{\nu_{4}}=135,\, m_{N_{4}}=2500$$\,$GeV
and the Higgs mass greater than 115$\,$GeV. In accordance with the
FD predictions, the $\nu_{4}-N_{4}$ mass ratio is selected to yield
a equivalent Dirac mass close to $m_{4}$ and the $u_{4}-d_{4}$ mass
difference is constrained to be less than $m_{W}$. Figure$~$\ref{fig:sin34m}
shows a scan of the Higgs boson mass from 150 up to 900$\,$GeV in
steps of 50$\,$GeV for 3 different values of the mixing angle between
the third and fourth generations. It can be seen that a larger mixing
angle increases T, compensating for larger Higgs boson mass values
pulling towards higher $S$ and lower $T$ values. The best cases
are obtained for either small mixing angle (up to $|\sin\theta_{34}|$
=0.05) and Higgs mass between 150 and 350 GeV. If the mixing angle
is increased up to $|\sin\theta_{34}|=0.1$, then a higher Higgs mass
range of 550-600 GeV becomes feasible. 

\begin{figure}[H]
\begin{centering}
\includegraphics[width=0.8\columnwidth]{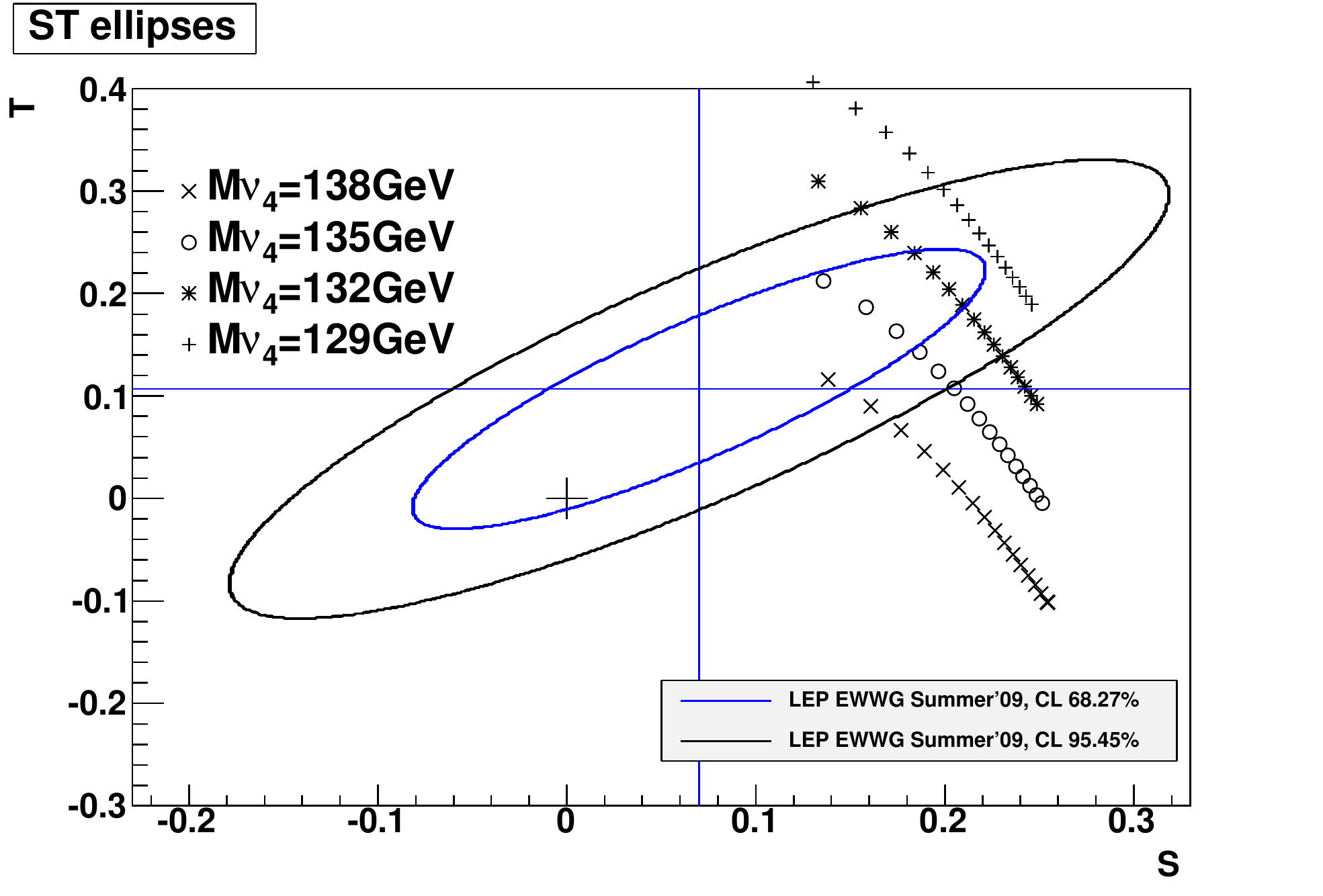}
\par\end{centering}

\caption{The Higgs mass dependence for various values of the fourth family
lighter neutrino mass, $\nu_{4}$ . }
\label{Fig:v4_H-majo}
\end{figure}

The Figure \ref{Fig:v4_H-majo} displays the dependency on the $\nu_{4}$
and Higgs mass values. The Higgs boson mass was scanned from 150 up
to 900 GeV in steps of 50$\,$GeV for 4 different $\nu_{4}$ masses:
129, 132, 135 and 138$\,$GeV. Other relevant parameters are $|\sin\theta_{34}|=0.02$,
$m{}_{u_{4}}$=545$\,$GeV, $m{}_{d_{4}}=m_{e_{4}}=550$$\,$GeV and
$m_{N_{4}}=2500$$\,$GeV. One can notice that the effect of increasing
the $\nu_{4}$ mass is the opposite of increasing the mixing angle,
$|\sin\theta_{34}|$ : it leads to smaller T values. Under the studied
conditions, the most compatible values are obtained for $m_{\nu_{4}}=135$$\,$GeV
and $150<m_{H}<300$$\,$GeV.

\begin{figure}[H]
\begin{centering}
\includegraphics[width=0.8\columnwidth]{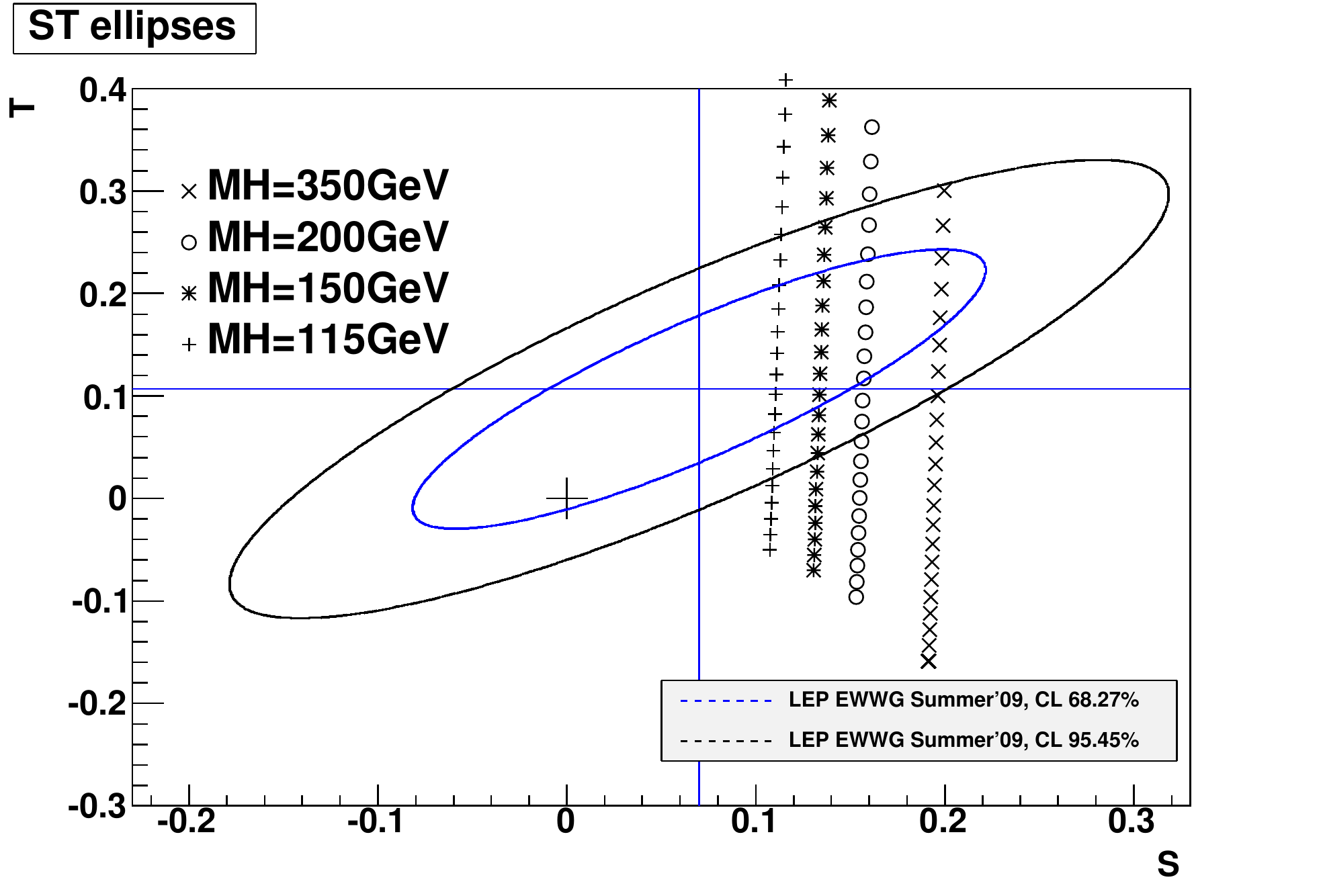}
\par\end{centering}

\caption{The fourth family heavier neutrino, $N_{4}$ mass dependence for various
values of the Higgs mass. }
\label{Fig:N4_H-majo}
\end{figure}

The Figure$~$\ref{Fig:N4_H-majo} displays the dependency on the
$N_{4}$ and Higgs mass values. The $N_{4}$ mass was scanned from
1900 up to 4000 GeV in steps of 100 GeV for four Higgs boson masses:
115, 150, 200 and 350$\,$GeV. The other relevant parameters are$|\sin\theta_{34}|=0.02$,
$m_{u_{4}}=545$$\,$GeV, $m_{d_{4}}=m_{e_{4}}=550$$\,$GeV and $m_{\nu_{4}}=135$$\,$GeV.
One can notice that the effect of increasing the $N_{4}$ mass is
to decrease $T$ sharply and $S$ very mildly. Once more the most
compatible results (closest to the EW precision fit minimum) are obtained
for the lightest possible Higgs mass (115$\,$GeV) for $N_{4}$ mass
between 2700 and 3300$\,$GeV but also the higher Higgs masses (e.g.
350$\,$GeV) stays within the $1\sigma$ error ellipse if the $N_{4}$
mass is between 2100 and 2300$\,$GeV.

\begin{figure}[H]
\begin{centering}
\includegraphics[width=0.8\columnwidth]{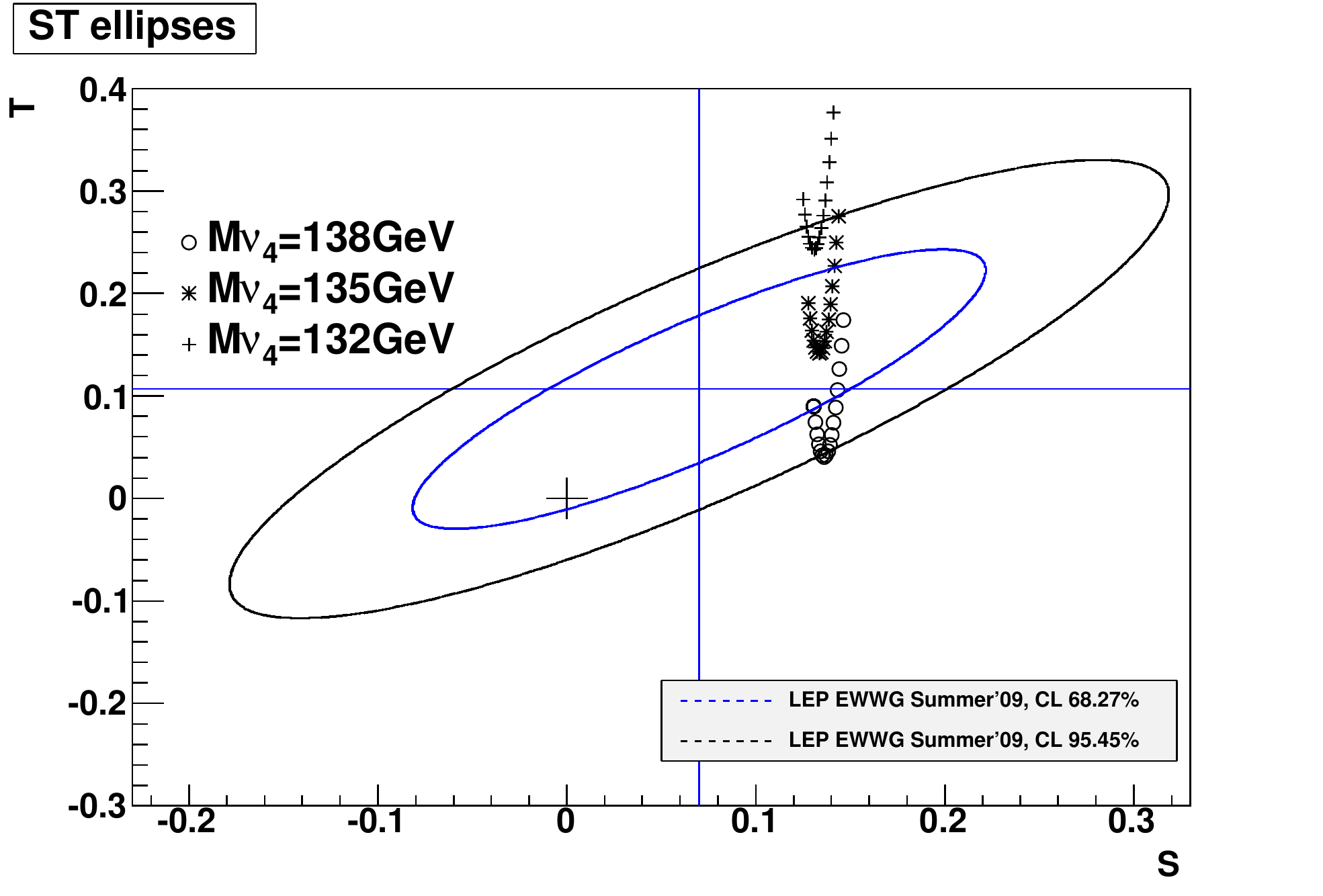}
\par\end{centering}

\caption{The fourth family up type quark, $u_{4}$ mass dependence for various
values of the fourth family lighter neutrino mass, $\nu_{4}$. }
\label{Fig:u4_v4-majo}
\end{figure}

Up to this point the $u_{4}-d_{4}$ mass difference was kept at 5$\,$GeV.
The impact of increasing this difference for different values of $\nu_{4}$
mass can be seen in Figure \ref{Fig:u4_v4-majo}. The $u_{4}$ mass
was scanned from 500 up to 580$\,$GeV in steps of 5$\,$GeV for three
values of $\nu_{4}$ mass: 132, 135, 138$\,$GeV. The other relevant
parameters are $m_{H}=150$$\,$GeV and $m_{N_{4}}=2800$$\,$GeV
(one of the best pairs found in the previous paragraph), $m{}_{d_{4}}=m_{e_{4}}=550$$\,$GeV
and $|\sin\theta_{34}|=0.02$. The $u_{4}$ mass increase leads to
an initial decrease of $T$ and after a minimum it results in an increase
in $T$ while the $S$ values keep decreasing very slightly. For example
the minimum for $m_{\nu_{4}}=135$$\,$GeV occurs around $m_{u_{4}}=545$$\,$GeV
but any value of $u_{4}$ mass between 530 and 560$\,$GeV is rather
close to the EW data fit minimum.

\begin{figure}[H]
\begin{centering}
\includegraphics[width=0.8\columnwidth]{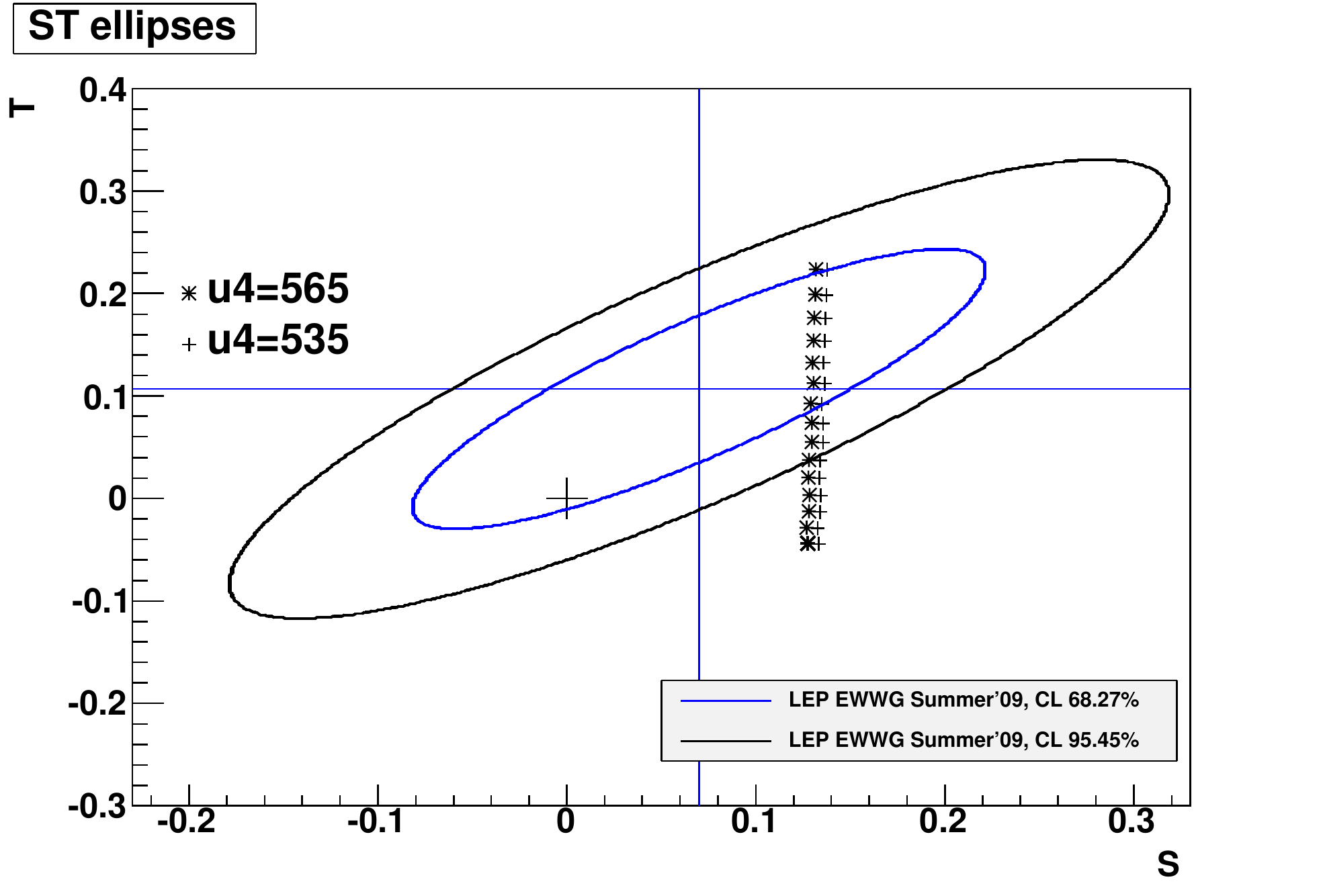}
\par\end{centering}

\caption{The fourth family up type quark, $u_{4}$ mass dependence for various
values of the fourth family heavier neutrino, $N_{4}$. }
\label{Fig:u4_N4-majo}
\end{figure}

To cross check the optimization of the heavier neutrinos mass, $m_{N_{4}}$,
it was also scanned from 2500 up to 4000$\,$GeV in steps of 100$\,$GeV
for two values of $u_{4}$ mass obtained in the previous paragraph.
The Figure$~$\ref{Fig:u4_N4-majo} shows the results of this scan.
The other relevant parameters are $|\sin\theta_{34}|=0.02$, $m{}_{\nu_{4}}=135$$\,$GeV,
$m{}_{d_{4}}=m_{e_{4}}=550$$\,$GeV and $m_{H}=150$$\,$GeV. It
can be seen that as in Figure$~$\ref{Fig:N4_H-majo}, the $N_{4}$
mass which yields the most compatible result is 2800$\,$GeV.

Finally, for different values of $m_{4}$, all the free parameters
of the Majorana case were scanned with the constraint $m_{Dirac}\equiv\sqrt{m{}_{N_{4}}\times m_{\nu_{4}}}=m_{4}$
while changing the \underbar{$m_{N_{4}}$ }to $m_{\nu_{4}}$ ratio.
The $u_{4}$ (Higgs) mass was scanned in steps of 5$\,$(10)$\,$GeV,
the mixing angle in steps of 0.01 and the $N_{4}$ to $\nu_{4}$ mass
ratio in steps of 2. The Table$~$\ref{Flo:Best-R-Majorana} contains
the mass and mixing parameters for the best $\triangle\chi^{2}$ values
obtained from a scan of all relevant parameters. It is seen that the
favoured values are $m_{H}=115$$\,$GeV, $|\sin\theta_{34}|=$0.02,
$m_{N_{4}}/m_{\nu_{4}}=14$ and $m_{u_{4}}-m_{4}=50$$\,$GeV, independent
of $m_{4}$ mass. We note that the minimum $\triangle\chi^{2}$ obtained
with the scan is much lower than the 3-family SM's $\triangle\chi^{2}=1.7$
for all three $m_{4}$ values. 

\begin{table}[H]
\caption{The fourth family parameters at the best $\triangle\chi^{2}$ values
compiled for FD hypothesis, Majorana scenario.}

\centering{}\begin{tabular}{c|c|c|c}
$m_{4}$ (GeV) & 400 & 550 & 700\tabularnewline
\hline 
$m_{u_{4}}$(GeV) & 450 & 600 & 750\tabularnewline
\hline 
$m_{N_{4}}$(GeV) & 1497 & 2058 & 2618\tabularnewline
\hline 
$m_{\nu_{4}}$(GeV) & 107 & 147 & 187\tabularnewline
\hline 
$|\sin\theta_{34}|$ & 0.02 & 0.02 & 0.02\tabularnewline
\hline 
$m_{H}$ (GeV) & 115 & 115 & 115\tabularnewline
\hline 
$\triangle\chi_{min}^{2}$ & 0.143 & 0.187 & 0.215\tabularnewline
\hline 
$S$ & 0.110 & 0.115 & 0.117\tabularnewline
\hline 
$T$ & 0.136 & 0.139 & 0.146\tabularnewline
\end{tabular}\label{Flo:Best-R-Majorana}
\end{table}

\section{Conclusions}

We have presented OPUCEM, an open-source function library publicly
available to calculate the electroweak oblique parameters $S,$ $T$
and $U$ for a number of physics models. This library has been heavily
tested and shown to be able reproduce various earlier results successfully.
It provides safe-to-use functions with an error-checking machinery
that relies on comparisons amongst different formulas and between
exact calculations and well-known approximations. The current version
is expected to compile, run and give correct results using any Unix-like
platform with the C++ Technical Report 1 (TR1) compiler support (e.g.
g++ 4.2 or later).

OPUCEM is meant to set an example to a new concept that we call {}``open-formula''
papers. Many scientific papers include numerical examples of various
symbolic calculations. With time, such examples loose a lot of their
value, unless they are not readily reproducible by others. Reproduction
of old results is usually very difficult, since most representations
of the numerical results are on plots and figures, which provide only
limited precision when one wants to check an independent re-implementation
of the symbolic calculations. As a solution to this, we suggest authors
of scientific papers, which present and demonstrate new formulas,
commit the actual computer code of their numerical calculations on
a publicly available repository. This will also help resolve errors
that occasionally happen when the results are sent to publication,
such as misprinting of the formulas.

In addition to the library itself, we have also presented an example
of its use in studying the mass parameters of a fourth generation
of Standard Model fermions. Investigations yielded regions of the
parameter space favored by the precision EW data, both for Dirac and
Majorana cases of the fourth SM family neutrinos. For the Dirac-type
neutrinos, the allowed mass parameters describe an elliptical ring
in the $m_{u_{4}}-m_{d_{4}}$, $m_{\nu_{4}}-m_{e_{4}}$ plane when
the sum of the fourth generation quark masses and the sum of the fourth
generation lepton masses are kept constant. The radius and the thickness
of this elliptical ring are dependent on the mass of the Higgs boson
and the mixing angle between the third and fourth generation quarks.
We also note that on this ring, EW data favours $m_{u_{4}}>m_{d_{4}}$
and $m_{\nu_{4}}<m_{e_{4}}$. For the Majorana case, the additional
parameter (the mass of the heavier neutrino) gives more flexibility
to the choice of parameters and especially by adjusting the ratio
of the masses of the new neutrinos, it is possible to move some additional
portions of the parameter space to become compatible with the measured
$S,\, T$ values. Finally, we test predictions of the Flavor Democracy
hypothesis and find that in either Majorana or Dirac case, for the
FD predictions to be true, the sine of the mixing angle should be
about 0.02, $m_{u_{4}}$ should be about 50-60 GeV higher than the
generic 4th generation mass scale and the smaller values of the Higgs
mass are preferred. Under these conditions, the $\Delta\chi^{2}$
measured on the $S,\, T$ plane for an FD-motivated 4-generation SM
is lower than the 3-generation SM.

\end{document}